\documentclass[10pt]{article}
\pdfoutput=1
\usepackage{jheparxiv}
\usepackage[latin1]{inputenc}
\usepackage{latexsym,diagbox,stackrel}
\usepackage{relsize}
\usepackage{shuffle}
\usepackage{amsmath}
\usepackage{mathrsfs}
\usepackage{array}
\usepackage{arydshln,leftidx,mathtools}
\usepackage{bbold}
\usepackage{graphicx}
\usepackage{subcaption}
\usepackage{pdflscape}
\usepackage{amssymb}
\usepackage{multirow}
\usepackage{enumerate}
\usepackage{blkarray}
\usepackage[usenames,dvipsnames,svgnames,table]{xcolor}
\usepackage{cleveref}
\usepackage{tikz}
\usepackage{graphicx}
\usepackage{slashed}
\usepackage{array}
\usepackage{pdfpages}
\usepackage[debug]{mciteplus}
\usetikzlibrary{shapes.geometric, arrows}
\definecolor{light-gray}{gray}{0.95}
\definecolor{light-grayII}{gray}{0.85}

\def\d{\text{d}}

\newcommand{\C}{\mathbb{C}}
\newcommand{\CP}{\mathbb{CP}}

\newcommand{\cI}{\mathcal{I}}

\newcommand{\Z}{\mathbb{Z}}

\newcommand{\SL}{\mathrm{SL}}

\newcommand{\tr}{\mathrm{tr}}

\newcommand{\pf}{\text{Pf} \,}

\newcommand{\px}{\raisebox{.15\baselineskip}{\Large\ensuremath{\wp}}}
\newcommand{\vol}{\mathrm{vol}\,}

\newcommand{\ra}{\mathrm{a}}

\subheader{\hfill \begin{tabular}{r} \texttt{QMUL-PH-19-18} \end{tabular}}
 
\title{\Large Two-Loop Scattering Amplitudes: Double-Forward Limit and Colour-Kinematics Duality}
\author[a]{Yvonne Geyer,}
\author[b]{ Ricardo Monteiro}
\author[b]{\& Ricardo Stark-Much\~ao}
        
\affiliation[a]{School of Natural Sciences, Institute for Advanced Study \\
        Einstein Drive, Princeton, NJ 08540, USA}

\affiliation[b]{Centre for Research in String Theory, School of Physics and Astronomy \\
        Queen Mary University of London, E1 4NS, United Kingdom}
        
\emailAdd{yvonnegeyer@ias.edu}
\emailAdd{ricardo.monteiro@qmul.ac.uk}
\emailAdd{r.j.stark-muchao@qmul.ac.uk}

\abstract{
We propose new formulae for the two-loop $n$-point $D$-dimensional integrands of scattering amplitudes in Yang-Mills theory and gravity. The loop integrands are written as a double-forward limit of tree-level trivalent diagrams, and are inferred from the formalism of the two-loop scattering equations. We discuss the relationship between the formulae for non-supersymmetric theories and the Neveu-Schwarz sector of the formulae for maximally supersymmetric theories, which can be derived from ambitwistor strings. An important property of the loop integrands is that they are expressed in a representation that includes linear-type propagators. This representation exhibits a loop-level version of the colour-kinematics duality, which follows directly from tree level via the double-forward limit. 
}

\begin{document}

\maketitle

\section{Introduction} \label{sec:intro}

Worldsheet techniques inspired by string theory offer an alternative to the Feynman diagram approach for calculating scattering amplitudes in quantum field theory, in particular for theories of massless particles. This broad programme has seen remarkable advances in recent years. Our aims are to extend the lessons learned at tree level and one loop to two-loop amplitudes, and to interpret previous two-loop results for maximally supersymmetric theories in a more general context, allowing for reduced or no supersymmetry. Even though we will motivate our proposal for two-loop amplitudes from the insights of this `stringy' approach, the proposal itself will not be written in a worldsheet language. It will instead be written in a (non-Feynman) diagrammatic language, whereby the two-loop integrands are suitably defined double-forward limits of tree-level amplitudes.

The worldsheet techniques that inspire our work originated in Witten's twistor string \cite{Witten:2003nn} describing four-dimensional super-Yang-Mills theory, and in the corresponding `connected prescription' to compute scattering amplitudes \cite{Roiban:2004yf}. In this approach, tree-level scattering amplitudes for $n$ massless particles are computed as integrals over the moduli space of punctured Riemann spheres, $\mathfrak{M}_{0,n}$. The modern version of these advances, applicable to theories of massless particles in any number of dimensions, was developed by Cachazo, He and Yuan (CHY) \cite{Cachazo:2013gna,Cachazo:2013hca,Cachazo:2013iea}, who discovered the general type of formulae, and by Mason and Skinner \cite{Mason:2013sva}, who constructed the associated type of worldsheet model; the ambitwistor string. For a variety of interesting theories in this framework, see e.g.~\cite{Cachazo:2014xea,Ohmori:2015sha,*Casali:2015vta}. The ambitwistor string models reproducing Yang-Mills and gravity amplitudes are supersymmetric, but the extraction of amplitudes in non-supersymmetric theories is possible even at loop level, as we shall discuss. For a comparison of the bosonic and supersymmetric ambitwistor strings, see \cite{Azevedo:2017lkz,*Azevedo:2017yjy,*Berkovits:2018jvm,Azevedo:2018dgo}. For recent work on moduli-space formulae tuned to a specific number of spacetime dimensions using the spinor-helicity framework, see \cite{Geyer:2014fka,*Heydeman:2017yww,*Cachazo:2018hqa,*Geyer:2018xgb,*Heydeman:2018dje,*Geyer:2019ayz,*Schwarz:2019aat}.

The extension of these ideas beyond tree level was initially developed by studying $g$-loop supergravity integrands as integrals over the moduli space of punctured genus-$g$ Riemann surfaces, $\mathfrak{M}_{g,n}$ \cite{Adamo:2013tsa,*Casali:2014hfa,Adamo:2015hoa}. The technical challenges of dealing with higher-genus surfaces motivated a simpler and more general formalism, based on nodal Riemann spheres, where the relevant moduli space is $\mathfrak{M}_{0,n+2g}$ for $g$ loops. It relates to the $\mathfrak{M}_{g,n}$-formalism via a residue theorem on the moduli space, when both formalisms are possible \cite{Geyer:2015bja,Geyer:2015jch,Geyer:2016wjx,Farrow:2017eol,Geyer:2017ela,Geyer:2018xwu}. In the nodal formalism, each node consists of a pair of punctures on the sphere, through which the states in that `loop' run. Here, we will be interested in the two-loop case, where the bi-nodal sphere pictured below is relevant.
\begin{figure}[h]
\begin{center}
\includegraphics[scale=1.5]{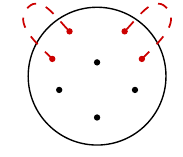} \\
\vspace{-.6cm}
\end{center}
\end{figure}

\noindent This line of work is closely related to conventional string theory approaches, where field theory amplitudes are obtained in the zero-slope limit.
As examples of recent work on two-loop field theory amplitudes from conventional string theory, see \cite{Magnea:2013lna,*Magnea:2015fsa,*DiVecchia:1996kf,*DiVecchia:2018dob} for the bosonic string, or \cite{Gomez:2015uha,Mafra:2015mja} for the superstring.

In this paper, we will pursue two important and closely related insights of worldsheet techniques, studied at one loop in recent years and now extended to two loops. One is a forward-limit construction of loop integrands, and the other is the loop-level manifestation of the colour-kinematics duality. Let us briefly discuss these.

The $n$-particle one-loop integrands obtained from the nodal sphere (one node) can be explicitly understood as a forward limit of an $(n+2)$-particle tree amplitude, with the node providing two punctures corresponding to the loop momentum, $\pm \ell_\mu$. This interpretation is made manifest in a new type of representation of the loop integrand, which includes non-Feynman propagators (whose inverse is linear, not quadratic, in the loop momentum) \cite{Geyer:2015bja,He:2015yua,Baadsgaard:2015twa,Geyer:2015jch,Cachazo:2015nwa,Roehrig:2017gbt}. It is not very surprising that such an interpretation of one-loop integrands exists, since this is the lesson of the Feynman tree theorem \cite{Feynman:1963ax,*Feynman:1972mt,*Feynman:2000fh}. At two loops, we will argue that a double-forward limit construction of two-loop integrands can be inferred from the worldsheet techniques, but is subtler for generic theories than for (maximally) supersymmetric ones. Generically, multiplicity coefficients are required for various diagrammatic contributions, reminiscent of those found in \cite{CaronHuot:2010zt} for planar theories (although we are not restricted to planarity here). It would be interesting to relate our construction to other recent approaches to the forward-limit in multi-loop amplitudes, e.g., ref.~\cite{Runkel:2019zbm}. 

The other insight we will pursue is the loop-level realisation of the `colour-kinematics duality' of Bern, Carrasco and Johansson (BCJ) \cite{Bern:2008qj}, which has been a major tool in the study of scattering amplitudes in gauge theory and gravity over the past decade. It states that it is possible to write a gauge-theory scattering amplitude in terms of trivalent diagrams, such that -- apart from the propagators -- the kinematic dependence mirrors the (Lie) algebraic structure of the colour dependence. Given this `duality,' gravity amplitudes are obtained by substituting the colour dependence with a second copy of the kinematic dependence. This `double copy' relating gauge theory and gravity is a diagrammatic version of the KLT relations known from string theory \cite{Kawai:1985xq,*Bern:1998ug}. Like the KLT relations, the colour-kinematics duality is associated to worldsheet monodromy relations, now in the field theory limit of string theory \cite{BjerrumBohr:2009rd,*Stieberger:2009hq}. Unlike the KLT relations, however, its diagrammatic interpretation allows for a natural conjecture of a loop-level extension, in particular for the loop integrand \cite{Bern:2010ue}. This technique, together with unitarity methods \cite{Bern:1994zx,*Bern:1994cg,*Britto:2004nc,*Ellis:2007qk}, has allowed for studies of the ultraviolet behaviour of supergravity theories.

At tree level, the colour-kinematics duality is well-established, even if its algebraic origin remains to be satisfactorily explained; see, for instance, the constructions \cite{Bern:2010yg,Tye:2010dd,Mafra:2011kj,Monteiro:2011pc,*BjerrumBohr:2012mg,*Monteiro:2013rya,Oxburgh:2012zr,Johansson:2014zca,*Johansson:2015oia,Chiodaroli:2014xia,*Chiodaroli:2015rdg,*Chiodaroli:2017ehv,*Chiodaroli:2018dbu,Anastasiou:2014qba,*Anastasiou:2017nsz,Naculich:2014naa,Lee:2015upy,*Bridges:2019siz,Brown:2016mrh,Bjerrum-Bohr:2016axv,Chen:2016fgi,Fu:2018hpu,Chen:2019ywi}. The status of the loop-level conjecture is unclear, however. At one loop, there are many examples, e.g. \cite{Boels:2011tp, *Boels:2011mn, Mafra:2012kh, Boels:2013bi, Bjerrum-Bohr:2013iza,  Bern:2013yya, Ochirov:2013xba, Mafra:2014gja, He:2015wgf, Primo:2016omk, Jurado:2017xut,Mafra:2018qqe}, but some potential obstructions have been found \cite{Mafra:2014gja, Berg:2016fui}. At higher loops, there is also plenty of supporting evidence for the conjectured duality, e.g. \cite{Bern:2010ue,Carrasco:2011mn,Bern:2012uf, *Bern:2012cd,*Bern:2013qca, *Bern:2013uka,*Bern:2014sna, Boels:2012ew,*Yang:2016ear, Mafra:2015mja,Johansson:2017bfl,*Kalin:2018thp,*Duhr:2019ywc,Anastasiou:2018rdx}, including the associated worldsheet monodromy relations \cite{Tourkine:2016bak,*Hohenegger:2017kqy,*Ochirov:2017jby,*Tourkine:2019ukp}. However, there are also known difficulties, most famously for maximal supersymmetry at four points, five loops. In this example, the difficulties were dealt with in \cite{Bern:2017ucb} by relaxing the strong constraint imposed by the colour-kinematic duality, following the general prescription of \cite{Bern:2017yxu} for evading the Jacobi relations. Previous work on avoiding difficulties dealt with the five-point two-loop amplitude in the absence of supersymmetry \cite{Bern:2015ooa,*Mogull:2015adi}, which is relevant for our present analysis. Naturally, there is a computational price to pay for relaxing the colour-kinematics duality, and it remains to be established what is the most appropriate loop-level extension of this very useful property of gauge theory and gravity.

We will build on work at one loop \cite{He:2016mzd,*He:2017spx,Geyer:2017ela}, where a different version of the loop-level colour-kinematics duality was described. This version is adapted to the type of representation of the loop integrand that we alluded to previously, which includes non-Feynman propagators. The advantage is that it easily takes tree-level relations into loop-integrand-level relations via the forward limit. We will discuss here the two-loop extension of this story. Translating these results into a standard representation of the loop integrand, directly comparable to the original colour-kinematics conjecture \cite{Bern:2010ue}, is an important question which we will not attempt to answer here.

Our main result is a proposal for the construction of two-loop integrands in Yang-Mills theory and gravity, with or without supersymmetry, via a double-forward limit. We sketch the proposal here, leaving the details for section~\ref{sec:proposal}. The $n$-point loop integrand can be expressed in terms of $(n+4)$-point trivalent tree-level diagrams as 
\begin{equation}
\quad \mathcal{A}^{(2)}_{\text{YM}} = \int \frac{\d^D\ell_1\,\d^D\ell_2}{\ell^2_1\,\ell^2_2} \sum_{\ra\,\in \Gamma^{(2)}_{n+4}} \frac{N^{(2)}_\ra\, c^{(2)}_\ra}{\rho_\ra\,D_\ra}\,,
 \qquad \;\;
  \mathcal{A}^{(2)}_{\text{grav}} = \int \frac{\d^D\ell_1\,\d^D\ell_2}{\ell^2_1\,\ell^2_2} \sum_{\ra\,\in \Gamma^{(2)}_{n+4}} \frac{N^{(2)}_\ra\, \tilde N^{(2)}_\ra}{\rho_\ra\,D_\ra
  \quad}\,.
\end{equation}
The colour factors $c_\ra^{(2)}$ are obtained from the tree-level ones by `gluing' the colour indices in each of the two nodes,
\begin{align}
c_\ra^{(2)} = \delta^{a_{1^+}a_{1^-}}\,\delta^{a_{2^+}a_{2^-}} c_\ra^{(0)} \,.
\end{align}
Likewise, the kinematic numerators are obtained from tree-level BCJ kinematic numerators,
\begin{align}
N_\ra^{(2)} = \sum_{r_1,r_2} N_\ra^{(0)} \,.
\end{align}
The sum is over polarisation states in the two nodes, and is implemented via completeness relations. The propagator sets $1/D_\ra$ are the ones appropriate for the loop-integrand representation used here. The numerical coefficients $\rho_\ra$ represent the multiplicity of standard two-loop diagrams in terms of the tree-level diagrams of our representation. Finally, the set of diagrams $\Gamma^{(2)}_{n+4}$ excludes those diagrams which are divergent in the double-forward limit. Notice that we do not sketch here versions of these formulae where the two-loop integrands are expressed as integrals over the moduli space $\mathfrak{M}_{0,n+4}$, analogous to the (maximally) supersymmetric formulae presented in \cite{Geyer:2018xwu}. It turns out that the picture of the bi-nodal sphere is not realised straightforwardly for generic theories.

This paper is organised as follows. In section~\ref{sec:review}, we review previous work at tree level, one loop and two loops, which motivates our construction, presented in section~\ref{sec:proposal}. In section~\ref{sec:NSsector}, we discuss the relation of our proposal to previous results obtained from ambitwistor string theory, in the supersymmetric case. We provide an illustration of our representation of the loop integrand in section~\ref{sec:checks}, with simple checks for a non-supersymmetric amplitude. Section~\ref{sec:discussion} is a brief discussion of the results and of future directions.

%%%%%%%%%%%%%%%%%%%%%%%%%%%%%%%%%%%%%%%%%%%%%%
%%%%%%%%%%%%%%%%%%%%%%%%%%%%%%%%%%%%%%%%%%%%%%

\vspace{.3cm}
\section{Review} \label{sec:review}

In this section, we start by reviewing the Cachazo-He-Yuan (CHY) formulae for tree-level scattering amplitudes in Yang-Mills theory and gravity, and their connection to the Bern-Carrasco-Johansson (BCJ) colour-kinematics duality. We then review the analogous construction for one-loop integrands, with an eye to its extension to two loops. Finally, we review the two-loop scattering equations.

\vspace{.3cm}
\subsection{Tree level}  \label{sec:reviewtree}

\vspace{.3cm}
\paragraph{Amplitudes from the scattering equations.} The CHY formalism \cite{Cachazo:2013gna,Cachazo:2013hca,Cachazo:2013iea} expresses tree-level scattering amplitudes for $n$ massless particles in $D$ dimensions as integrals over the moduli space of a punctured Riemann sphere, $\mathfrak{M}_{0,n}$. For null external momenta $k_i$, with $i=1,\ldots,n$, the amplitudes are written as
\begin{equation}
 {\mathcal{A}}^{(0)}_n=\int_{\mathfrak{M}_{0,n}}\!\!\d\mu_{n}^{(0)}\,\cI^{(0)}\,,\qquad\qquad \d\mu_{n}^{(0)}\equiv \frac{\d^n\sigma}{\vol\SL(2,\C)} \ \prod_i{}' \, \delta\Bigg(\sum_{j\neq i} \frac{k_i\cdot k_j}{\sigma_{ij}}\Bigg)\,,\label{eq:tree-ampl}
\end{equation}
with $\sigma_i\in\CP^1$ and $\sigma_{ij}=\sigma_i-\sigma_j$. We use the superscript ${}^{(0)}$ to denote tree level. The formalism is applicable to any theory of massless particles, and the measure $\d\mu_{n}^{(0)}$ is universal.\footnote{`Universal' here means that any massless theory can be written in the form \eqref{eq:tree-ampl}, where the integrand $\cI^{(0)}$ only depends on the kinematic data polynomially.}

Let us first consider the measure in detail. The  moduli space $\mathfrak{M}_{0,n}$ is parametrised by $\{\sigma_1,\cdots,\sigma_n\}$ up to an $\SL(2,\C)$ transformation of the $\sigma_i$, so that $\mathrm{dim}_{\mathbb C}(\mathfrak{M}_{0,n})=n-3$. An explicit expression for the measure is given by fixing three of the punctures, say for $i=r,s,t$, and excluding three (non-independent) delta functions, say for $i=r',s',t'$, so that
\begin{equation}
\d\mu_{n}^{(0)} = \sigma_{rs}\sigma_{st}\sigma_{tr}\Bigg( \prod_{i\neq r,s,t} \d\sigma_i \Bigg)  
\sigma_{r's'}\sigma_{s't'}\sigma_{t'r'} 
\prod_{i\neq r',s',t'} \delta\Bigg(\sum_{j\neq i} \frac{k_i\cdot k_j}{\sigma_{ij}}\Bigg)  \,.
\end{equation}
The measure therefore fully localises the moduli space integral onto the solutions of the so-called {\it scattering equations},
\begin{equation}
 E_i^{(0)}\equiv\sum_{j\neq i} \frac{k_i\cdot k_j}{\sigma_{ij}}=0\,.
\label{eq:SE}
\end{equation}
Only $n-3$ of the equations are linearly independent, and there exist $(n-3)!$ solutions up to $\SL(2,\C)$ transformations. 
The scattering equations can be associated to a null condition,
\begin{equation}
P^2(\sigma)=0 \,, \qquad \text{where} \qquad
P^\mu(\sigma)=d\sigma\sum_{i=1}^n \frac{k_i^\mu}{\sigma-\sigma_i}\,.
\label{eq:P2}
\end{equation}
In particular, $P^2$ vanishes on $\CP^1$ if and only if it has no poles, and $\text{Res}_{\sigma_i} P^2 = d\sigma_i \,2E_i^{(0)}$\,.

While the CHY measure is universal, the CHY integrand $\cI^{(0)}$ is theory-specific; see \cite{Cachazo:2014xea} for various examples. Here, we are mostly interested in the original CHY examples of Yang-Mills theory and gravity,
\begin{align}
 \cI_{\text{YM}}^{(0)}=\cI_{\text{kin}}^{(0)}\, \cI_{\text{SU}(N_c)}^{(0)}\,,\qquad\qquad \cI_{\text{grav}}^{(0)}=\cI_{\text{kin}}^{(0)}\,\widetilde\cI_{\text{kin}}^{(0)}\,,
 \label{eq:CHY_YM_grav}
\end{align}
for which we need two building blocks, one related to the colour dependence, $\cI_{\text{SU}(N_c)}^{(0)}$, and another related to the external kinematic data, $\cI_{\text{kin}}^{(0)}$. The `Parke-Taylor factor' carrying the Lie algebra structure is
\begin{align}
\cI_{\text{SU}(N_c)}^{(0)}(\{a_i,\sigma_i\})=\sum_{\rho\in {S}_{n}/\Z_n}\frac{\tr\left(T^{\rho(a_1)}T^{\rho(a_2)}\cdots T^{\rho(a_n)}\right)}{\sigma_{\rho(a_1)\rho(a_2)}{\sigma_{\rho(a_2)\rho(a_3)}\cdots{\sigma_{\rho(a_n)\rho(a_1)}}}}\,,\label{equ:Icoltreetr}
\end{align}
where the $a_i$ are the Lie algebra indices of the external gluons, and the sum is over non-cyclic permutations. The kinematic factor is the `CHY reduced Pfaffian',
\begin{align}
\cI_{\text{kin}}^{(0)}(\{\epsilon_i,k_i,\sigma_i\})=\pf'(M)\equiv\frac{(-1)^{\hat i+\hat j}}{\sigma_{\hat i\hat j}}\pf\left(M^{\hat i \hat j}_{\hat i \hat j}\right)\,.% \qquad \widetilde\cI_{\text{kin}}=\cI_{\text{kin}}(\epsilon_i\to\tilde\epsilon_i)\,,
\label{equ:Ikintreepf}
\end{align}
Here, M is a $2n\times 2n$ antisymmetric matrix defined in terms of the momenta $k_i$, the polarisation vectors $\epsilon_i$ and the marked points $\sigma_i$ as
 \begin{subequations}\label{eq2:def_M_CHY}
 \begin{align} 
  & &&M=\begin{pmatrix} A & -C^T \\ C & B \end{pmatrix}\,, &&\\
  &A_{ij}=\frac{k_i\cdot k_j}{\sigma_{ij}}\,,&&B_{ij}=\frac{\epsilon_i\cdot \epsilon_j}{\sigma_{ij}}\,, &&C_{ij}=\frac{\epsilon_i\cdot k_j}{\sigma_{ij}}\,,\\
  &A_{ii}=0\,, && B_{ii}=0\,, && C_{ii}=-\sum_{j\neq i}C_{ij}\,.
 \end{align}
 \end{subequations}
If the $\sigma_i$ satisfy the scattering equations \eqref{eq:SE}, then $M$ has co-rank two,\footnote{Its kernel is spanned by the vectors $(1,\cdots,1,0,\cdots,0)$ and $(\sigma_1,\cdots,\sigma_n,0,\cdots,0)$.} and therefore $\pf(M)=0$. The reduced Pfaffian $\pf'(M)$ is defined by removing any two rows and columns $\hat i$ and $\hat j$ such that $1\leq \hat i < \hat j \leq n$. If the $\sigma_i$ satisfy the scattering equations, then $\pf'(M)$ is independent of the choice of $\hat i$ and $\hat j$. Remarkably, it is also gauge invariant, so that this important property is manifest in the CHY formulae.

The CHY integrand for gravity, $\cI_{\text{grav}}^{(0)}$, has a factor $\cI_{\text{kin}}^{(0)}$ and a factor $\widetilde\cI_{\text{kin}}^{(0)}=\cI_{\text{kin}}^{(0)}(\epsilon_i\to\tilde\epsilon_i)$. The sets of polarisation vectors ${\epsilon_i^\mu}$ and $\tilde\epsilon_i^\mu$ make up the set of polarisation tensors $\varepsilon_i^{\mu\nu}=\epsilon_i^\mu\tilde\epsilon_i^\nu$ of the external states. States of this type form a basis for the (generically non-factorisable) states of NS-NS gravity, describing Einstein gravity coupled to a dilaton and a (2-form) B-field.\footnote{The name NS-NS gravity originates in string theory, where this is the low-energy effective theory for the massless level of the closed string -- in particular, when both left and right movers are in the Neveu-Schwarz (NS) sector.} The restriction to pure Einstein gravity, i.e., only gravitons, is achieved by choosing appropriate symmetric and traceless polarisation tensors. At loop level, the additional states will also run in the loops, but we will discuss below how they can be projected out.

\vspace{.3cm}
\paragraph{Trivalent diagrams and colour-kinematics duality.} The elegant symmetry exhibited in \eqref{eq:CHY_YM_grav} between colour and kinematics, and between Yang-Mills theory and gravity, is reminiscent of the BCJ colour-kinematics duality \cite{Bern:2008qj}, and also of the older Kawai-Lewellen-Tye relations \cite{Kawai:1985xq,*Bern:1998ug}. We will see in the following how to relate the CHY formulae to the colour-kinematics duality.

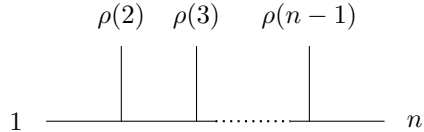
\begin{figure}[ht]
\begin{center}
 \begin{tikzpicture}[scale=1]
 \draw (0,0) -- (2.25,0) ;
 \draw[dotted, thick] (2.25,0) -- (3.25,0) ;
 \draw (3.25,0) -- (4.5,0);
 \draw (1,0) -- (1,1) ;
 \draw (2,0) -- (2,1) ;
 %\draw (3,0) -- (3,1) ;
 \draw (3.5,0) -- (3.5,1) ;
 \node at (-0.4,0) {$1$};
 \node at (4.9,0) {$n$};
 \node at (1,1.4) {$\rho(2)$};
 \node at (2,1.4) {$\rho(3)$};
 %\node at (2.75,0.5,0) {$...$};
 \node at (3.5,1.4) {$\rho(n-1)$};
\end{tikzpicture}
\end{center}
\caption{The tree-level BCJ master diagrams are half-ladder diagrams with fixed endpoints, which we choose to be legs $1$ and $n$.}
\label{fig:half-ladder}
\end{figure}

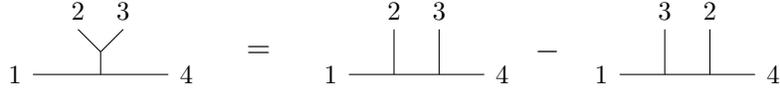
\begin{figure}[ht]
\begin{center}
 \begin{tikzpicture}[scale=0.6]
 \draw (0,0) -- (3,0) ;
 \draw (1,0) -- (1,1) ;
 \draw (2,0) -- (2,1) ;
 \node at (-0.4,0) {$1$};
 \node at (3.4,0) {$4$};
 \node at (1,1.4) {$2$};
 \node at (2,1.4) {$3$};
 \node at (4.4,0.5) {{\large $-$}};
 \draw (6,0) -- (9,0) ;
 \draw (7,0) -- (7,1) ;
 \draw (8,0) -- (8,1) ;
 \node at (5.6,0) {$1$};
 \node at (9.4,0) {$4$};
 \node at (7,1.4) {$3$};
 \node at (8,1.4) {$2$};
 \node at (-2,0.5) {{\large $=$}};
 \draw (-7,0) -- (-4,0) ;
 \draw (-5.5,0) -- (-5.5,0.5) ;
 \draw (-5.5,0.5) -- (-5,1) ;
 \draw (-5.5,0.5) -- (-6,1) ;
 \node at (-7.4,0) {$1$};
 \node at (-3.6,0) {$4$};
 \node at (-6,1.4) {$2$};
 \node at (-5,1.4) {$3$};
\end{tikzpicture}
\end{center}
\caption{Example at 4 points of how a diagram relates to master diagrams via Jacobi relations. This equation applies to the colour factors of the diagrams. We shall see that it applies also to their kinematic BCJ numerators.}
\label{fig:jacobiexample}
\end{figure}

There is a natural procedure to translate CHY formulae into trivalent Feynman-like diagrams, which also allows us to avoid solving the scattering equations explicitly. To start with, we can use an old result \cite{DelDuca:1999rs} to write
\begin{align}
\cI_{\text{SU}(N_c)}^{(0)} \; = \sum_{\rho\in S_{n-2}} 
\frac{ c\big(1,\rho(2,\cdots,n-1),n\big) 
}{\sigma_{1\rho(2)}\,\sigma_{\rho(2)\rho(3)} \cdots \sigma_{\rho(n-1)n} \,\sigma_{n1}}\,,\label{equ:Icoltreecf}
\end{align}
where instead of the colour traces of \eqref{equ:Icoltreecf} we employ colour factors,
$$
c\big(1,\rho(2,\cdots,n-1),n\big)  = f^{a_1a_{\rho(2)}b_1} \, f^{b_1a_{\rho(3)}b_2} \cdots f^{b_{n-3}a_{\rho(n-1)}a_n}\,, \qquad  \text{with} \quad [T^a,T^b]=f^{abc} T^c \,.
$$
These colour factors are associated to the `half-ladder' diagrams pictured in \cref{fig:half-ladder}. They can be taken as the {\it master diagrams}, since the colour factor of any other trivalent diagram can be obtained from the colour factors of the master diagrams using the Jacobi identity for the Lie algebra of $\text{SU}(N_c)$. We give a basic example in \cref{fig:jacobiexample}.

There is an analogous decomposition of the CHY kinematic factor \cite{Cachazo:2013iea}.\footnote{This decomposition follows from the property of `KLT orthogonality' \cite{Cachazo:2013gna, Cachazo:2012da}. An entirely analogous decomposition was previously found in open superstring amplitudes in the pure spinor formalism \cite{Mafra:2011nv}, and the relation to the colour-kinematics duality was presented in \cite{Mafra:2011kj}. See \cite{Mizera:2017sen,*Mizera:2017rqa,Azevedo:2018dgo,He:2018pol,*He:2019drm} for some recent work on this connection between moduli-space integrands in string theory and in field theory.} The reduced Pfaffian \eqref{equ:Ikintreepf} can be re-expressed as follows if the $\sigma_i$ satisfy the scattering equations,
\begin{align}
E_i^{(0)}=0\quad \Rightarrow \quad \cI_{\text{kin}}^{(0)}\;=\sum_{\rho\in S_{n-2}} 
\frac{ N\big(1,\rho(2,\cdots,n-1),n\big)
}{\sigma_{1\rho(2)}\,\sigma_{\rho(2)\rho(3)} \cdots \sigma_{\rho(n-1)n} \,\sigma_{n1}} \,.
\label{equ:Ikintreept}
\end{align}
The kinematic numerators $N(\cdot)$ depend only on the Lorentz invariants $k_i\cdot k_j$, $\epsilon_i\cdot k_j$ and $\epsilon_i\cdot \epsilon_j$, i.e., not on the $\sigma_i$. The numerators appearing in this formula are associated to master diagrams, as in \cref{fig:half-ladder}. Numerators for all other trivalent diagrams are defined via Jacobi-type relations that precisely mirror the Jacobi identities involving colour factors. In the basic example of \cref{fig:jacobiexample}, the equation applies to the kinematic numerators by construction, as well as to the colour factors. Notice that the decomposition \eqref{equ:Ikintreept} is non-unique. Of particular interest are sets of local numerators, i.e., numerators that are polynomial in the Lorentz invariants. A useful explicit algorithm to obtain sets of local numerators from the reduced Pfaffian was presented in \cite{Fu:2017uzt,*Du:2017kpo}.

With colour factors and kinematic numerators for all trivalent diagrams, the Yang-Mills amplitude can be written as
\begin{equation}
\label{equ:cubicYMtree}
 \mathcal{A}^{(0)}_{\text{YM}}=\int_{\mathfrak{M}_{0,n}}\!\!\d\mu_{n}^{(0)}\;\cI_{\text{kin}}^{(0)}\; \cI_{\text{SU}(N_c)}^{(0)}
 = \sum_{\ra\in \Gamma_n} \frac{N_\ra\, c_\ra}{D_\ra}\,,
\end{equation}
where $\Gamma_n$ denotes the set of trivalent $n$-point diagrams. The $1/D_\ra$ denote scalar propagator factors for each diagram, that is, $1/K^2$ per internal leg if $K$ is the momentum flowing through it. We stress that the kinematic numerators defined as above satisfy the same algebraic relations as the colour factors, in particular the same Jacobi-type relations,
\begin{equation}
 c_\ra \pm c_{\mathrm{b}} \pm c_{\mathrm{c}}=0 \qquad \longleftrightarrow \qquad
 N_\ra \pm N_{\mathrm{b}} \pm N_{\mathrm{c}}=0\,. \label{equ:BCJjacobi}
\end{equation}
The fact that such a representation of the Yang-Mills amplitude is possible is known as the {\it colour-kinematics duality} or {\it BCJ duality} \cite{Bern:2008qj}. Kinematic numerators satisfying the same algebraic properties as the colour factors, in particular \eqref{equ:BCJjacobi}, are known as {\it BCJ numerators}.\footnote{Notice that the usual Feynman rules do not lead to trivalent diagrams only, due to the four-point vertex. If all the contributions obtained from the Feynman rules are grouped by colour factors, which are intrinsically trivalent, then one can trivially assign kinematic numerators to trivalent diagrams, but these will generically not satisfy the colour-kinematics duality, except at four points.} With these numerators in hand, we can immediately construct gravity amplitudes,
\begin{equation}
\label{equ:cubicgravtree}
 \mathcal{A}^{(0)}_{\text{grav}}=\int_{\mathfrak{M}_{0,n}}\!\!\d\mu_{n}^{(0)}\;\cI_{\text{kin}}^{(0)}\; \widetilde\cI_{\text{kin}}^{(0)}
 = \sum_{\ra\in \Gamma_n} \frac{N_\ra\, \widetilde N_\ra}{D_\ra}\,,
\end{equation}
where $\widetilde N=N(\epsilon_i\to\tilde\epsilon_i)$. The expressions \eqref{equ:cubicYMtree} and \eqref{equ:cubicgravtree} follow from the fact that the moduli space integrals
$$
\int_{\mathfrak{M}_{0,n}}\!\!\d\mu_{0,n}\;
\frac{1}{\sigma_{\rho(1)\rho(2)}\cdots\sigma_{\rho(n-1)\rho(n)}\sigma_{\rho(n)\rho(1)}}\;
\frac{1}{\sigma_{\rho'(1)\rho'(2)}\cdots\sigma_{\rho'(n-1)\rho'(n)}\sigma_{\rho'(n)\rho'(1)}}\,,
$$
where $\rho,\rho'\in S_n$, give rise to linear combinations of propagator sets $1/D_\ra$ with coefficients $\pm 1$ \cite{Cachazo:2013iea}. These linear combinations are precisely such that the colour factors and kinematic numerators of non-master diagrams are determined via Jacobi relations from those of the master diagrams in \eqref{equ:Icoltreecf} and \eqref{equ:Ikintreept}, respectively.

%%%%%%%%%%%%%%%%%%%%%%%%%%%%%%%%%%%%%%%%%%%%%%

\vspace{.3cm}
\subsection{One loop}  \label{sec:review1loop}

\vspace{.3cm}
\paragraph{Loop integrands from the scattering equations.} The known loop-level extension of the CHY formalism applies to the loop integrand. It was originally developed from ambitwistor string theory, a worldsheet model of field theory proposed in \cite{Mason:2013sva} that leads naturally to CHY formulae for scattering amplitudes at tree level. Ambitwistor string theory gives a prescription for computing one-loop amplitudes in certain theories by considering `genus-one scattering equations' \cite{Adamo:2013tsa,*Casali:2014hfa}. A generic and significantly simpler formalism was found in \cite{Geyer:2015bja,Geyer:2015jch} by applying the residue theorem to the genus-one moduli space, localising on a degenerate limit (where the modular parameter becomes $\tau=i\infty$) and thereby obtaining formulae on the punctured Riemann sphere (moduli space ${\mathfrak{M}_{0,n+2}}$), rather than on the torus (moduli space ${\mathfrak{M}_{1,n}}$). The formulae for the loop integrands are similar to the ones for tree-level amplitudes, but two extra punctures arise representing loop momentum insertions $\pm \ell$, as in the figure below. \\
\vspace{-.3cm}
\begin{figure}[h]
\begin{center}
\includegraphics[scale=2]{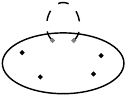}\\
\vspace{-.5cm}
\end{center}
\caption{Sphere with a single node.}
\label{nodalsphere}
\end{figure}
\vspace{.1cm}\\
For Yang-Mills theory, we have
\begin{equation}
\label{equ:seYM1loop}
 \mathcal{A}^{(1)}_{\text{YM}}=\int \frac{\d^D\ell}{\ell^2} \int_{\mathfrak{M}_{0,n+2}}\!\!\d\mu_{n+2}^{(1)}\;\cI_{\text{kin}}^{(1)}\; \cI_{\text{SU}(N_c)}^{(1)} \,,
\end{equation}
while for gravity, we have
\begin{equation}
\label{equ:segrav1loop}
 \mathcal{A}^{(1)}_{\text{grav}}=\int \frac{\d^D\ell}{\ell^2} \int_{\mathfrak{M}_{0,n+2}}\!\!\d\mu_{n+2}^{(1)}\;\cI_{\text{kin}}^{(1)}\; \widetilde\cI_{\text{kin}}^{(1)} \,.
\end{equation}

The CHY-type integrals in the loop integrands are associated to $n+2$ punctures with momentum insertions, of which $n$ relate to external momenta $\{k_i,\sigma_i\}$ and two relate to the loop momentum $\{\pm\ell,\sigma_\pm\}$. The various $n$-point ingredients can be defined in terms of their tree-level $(n+2)$-point counterparts by considering a null version of the loop momentum:\footnote{One may think of the null version $L$ as a higher-dimensional extension of $\ell$. This is analogous to the consideration of a pair of back-to-back massive momenta in the tree-level scattering equations \cite{Naculich:2014naa}. However, the completeness relation \eqref{equ:1loopcompleteness}, which will effectively implement the massless-vector state sum, is still a projector into a $(D-2)$-dimensional space.}
\begin{equation}
\label{equ:1loopL}
%\ell \leadsto 
L \qquad \textrm{such that} \qquad L^2=0\,, \quad L\cdot k_i \,\mapsto\, \ell\cdot k_i \,, \quad L\cdot \epsilon_i \,\mapsto\, \ell\cdot \epsilon_i\,.
\end{equation}
The measure is therefore defined as
\begin{equation}
\label{equ:1loopSEmeasure}
\d\mu_{n+2}^{(1)}\equiv \frac{\d^{n+2}\sigma}{\vol\SL(2,\C)} \ \prod_a{}' \, \delta\left(E_a^{(1)}\right)\,,
\end{equation}
where $a=\{i,\pm\}$, and the one-loop scattering equations read
\begin{equation}
\label{equ:1loopSE}
E_i^{(1)}=\frac{k_i\cdot \ell}{\sigma_{i+}}-\frac{k_i\cdot \ell}{\sigma_{i-}} +\sum_{j\neq i} \frac{k_i\cdot k_j}{\sigma_{ij}}\,, \qquad \qquad
E_\pm^{(1)}= \sum_{i} \frac{\pm\ell\cdot k_i}{\sigma_{\pm i}}\,.
\end{equation}
Notice that they do not depend on $\ell^2$, and indeed the tree-level-like interpretation relies on a null momentum $L$. As in \eqref{eq:P2}, the one-loop scattering equations are associated to the vanishing of a quadratic differential,
\begin{equation}
P^2(\sigma)-\ell^2\omega(\sigma)^2=0 \,, \quad \text{where} \quad
P^\mu(\sigma)=\ell^\mu\omega(\sigma)+\sum_{i=1}^n \frac{k_i^\mu}{\sigma-\sigma_i}\,,
\quad
\omega(\sigma) = \frac{\sigma_{+-}}{(\sigma-\sigma_+)(\sigma-\sigma_-)}\,.
\label{eq:P2_1loop}
\end{equation}
This corresponds to $P^2(\sigma)|_{\ell^2=0}=0$, that is, to the use of $\ell\rightarrow L$ from \eqref{equ:1loopL}.

Ref.~\cite{He:2015yua} studied in detail this forward limit from $(n+2)$-point tree level to $n$-point one loop at the level of the solutions to the scattering equations, with the following conclusion. The original $(n-1)!$ solutions from tree level (for $n+2$ particles with generic kinematics) split into three sets in the forward limit: set (i) with $(n-1)!-2(n-2)!$ regular solutions, where all $\sigma_a$ are distinct, that contribute at one loop; set (ii) with $(n-2)!$ singular solutions, where $\sigma_+=\sigma_-$, that contribute at one loop; and set (iii) with the remaining $(n-2)!$ singular solutions, where $\sigma_+\to\sigma_-$ in the forward limit, that are not solutions to the one-loop scattering equations, and do not contribute at one loop. In fact, it was shown in \cite{Geyer:2015jch,Cachazo:2015aol} that, while the set (ii) of solutions contributes to the loop integrand for generic theories, this contribution vanishes upon loop integration, so that the amplitude can be fully determined from the set (i) of regular solutions. At the level of the loop integrand, the separate contributions from the sets (i) and (ii) may give rise (depending on the theory) to cumbersome terms with discriminant poles, which vanish upon loop integration \cite{Cachazo:2015aol}; these terms cancel in the loop integrand between the sets (i) and (ii). The case of theories with maximal supersymmetry (or at least half-maximal for supergravity) is straightforward in this respect, as only the set (i) contributes to the loop integrand.

The remaining ingredients of the one-loop formulae \eqref{equ:seYM1loop} and \eqref{equ:segrav1loop} can also be defined in terms of the forward limit. For the one-loop colour factor, we simply glue the colour indices of the loop punctures,
\begin{equation}
\label{equ:Icol1loopfl}
\cI_{\text{SU}(N_c)}^{(1)} (\{a_i,\sigma_i\},\{\sigma_\pm\}) = \delta^{a_+a_-}\, \cI_{\text{SU}(N_c)}^{(0)} (\{a_i,\sigma_i\},\{a_\pm,\sigma_\pm\}) \,,
\end{equation}
where we consider the normalisation $\tr(T^aT^b)=\delta^{ab}$. The one-loop kinematic factor is 
\begin{equation}
\label{equ:Ikin1looppf}
\cI_{\text{kin}}^{(1)} (\{\epsilon_i,k_i,\sigma_i\},\{\ell,\sigma_\pm\}) = \sum_r\, \cI_{\text{kin}}^{(0)} (\{\epsilon_i,k_i,\sigma_i\},\{\epsilon_\pm^{r},\pm L,\sigma_\pm\}) \,,
\end{equation}
in the non-supersymmetric case \cite{Geyer:2015jch}.
The sum over the states running in the loop is defined by the state projector 
\begin{equation}
\label{equ:1loopcompleteness}
\sum_r\, \epsilon_{+\,\mu}^{\,r}\, \epsilon_{-\,\nu}^{\,r} = \eta_{\mu\nu} - \frac{L_\mu q_\nu+L_\nu q_\mu}{L\cdot q} \equiv \Delta_{\mu\nu} \,, \qquad
\text{such that} \quad \Delta^\mu_{\mu}=D-2\,,
\end{equation}
with a null reference vector $q_\mu$ \cite{Roehrig:2017gbt}. We can therefore write, in terms of the CHY Pfaffian,
\begin{equation}
\label{equ:Ikin1looppfM}
\cI_{\text{kin}}^{(1)} = \sum_r\, \pf'(M) = \Delta_{\mu\nu}\, \pf'(M)^{\mu\nu}\,,
\end{equation}
where $M$ is now a $2(n+2)\times2(n+2)$ matrix. In this construction, the loop integrand is gauge invariant (including for the choice of $q$) if only regular solutions to the scattering equations are considered, for the same reason that gauge invariance holds for the tree-level CHY Pfaffian. If singular solutions are also considered, then the loop integrand is not gauge invariant, but the amplitude is, as proven in \cite{Cachazo:2015aol}. Given the independence on the choice of $q$, we can use an effective substitution rule that avoids its use altogether \cite{Roehrig:2017gbt}:
\begin{equation}
\label{equ:1loopcompletenessrule}
\Delta_{\mu\nu} V^\mu W^\nu \leadsto V\cdot W \,,
\qquad \text{for any} \quad V,W\in\{k_i,\epsilon_i\}\,,
\end{equation}
whereas $\Delta_{\mu\nu}\, L^\nu=0$\,.

The supersymmetric counterpart of the one-loop kinematic factor \eqref{equ:Ikin1looppf} was derived from ambitwistor string theory \cite{Geyer:2015bja}. With hindsight, the forward limit described above is a natural result, but it was originally recognised by rewriting the Neveu-Schwarz contribution to the supersymmetric formula \cite{Geyer:2015jch}.

Let us make some comments regarding gravity. In the non-supersymmetric case, since both kinematic factors in  \eqref{equ:segrav1loop} are given by \eqref{equ:Ikin1looppf} -- one of them with `tilded' polarisations -- the relevant theory is NS-NS gravity, with graviton, dilaton and B-field, as discussed at tree level. If both of them were instead the supersymmetric kinematic factors of \cite{Geyer:2015bja}, then the relevant theory is maximal supergravity. If only one of the factors is the supersymmetric version, then we have half-maximal supergravity. Finally, returning to the non-supersymmetric case, notice that we can also consider pure Einstein gravity. Along with a choice of external polarisation tensors corresponding to gravitons, we should also project the dilaton and the B-field out of the loop by a judicious choice of the state projector
\begin{equation}
\sum_{r,\tilde r}\, \epsilon_{+\,\mu}^{\,r}\, \epsilon_{-\,\nu}^{\,r} \, \tilde\epsilon_{+\,\tilde\mu}^{\,\tilde r}\, \tilde\epsilon_{-\,\tilde\nu}^{\,\tilde r} \,,
\end{equation}
distinct from the NS-NS case $\Delta_{\mu\nu}\Delta_{\tilde\mu\tilde\nu}$. The pure Einstein case is \begin{equation}
\frac1{2}(\Delta_{\mu\nu}\Delta_{\tilde\mu\tilde\nu}+\Delta_{\mu\tilde\nu}\Delta_{\tilde\mu\nu}) - \frac1{D-2}\, \Delta_{\mu\tilde\mu}\Delta_{\nu\tilde\nu} \,,
\end{equation}
where the index symmetrisation eliminates the B-field, and the last term eliminates the dilaton.

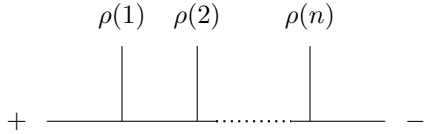
\begin{figure}[t]
\begin{center}
 \begin{tikzpicture}[scale=1]
 \draw (0,0) -- (2.25,0) ;
 \draw[dotted, thick] (2.25,0) -- (3.25,0) ;
 \draw (3.25,0) -- (4.5,0);
 \draw (1,0) -- (1,1) ;
 \draw (2,0) -- (2,1) ;
 %\draw (3,0) -- (3,1) ;
 \draw (3.5,0) -- (3.5,1) ;
 \node at (-0.4,0) {$+$};
 \node at (4.9,0) {$-$};
 \node at (1,1.4) {$\rho(1)$};
 \node at (2,1.4) {$\rho(2)$};
 %\node at (2.75,0.5,0) {$...$};
 \node at (3.5,1.4) {$\rho(n)$};
\end{tikzpicture}
\end{center}
\caption{The one-loop BCJ master diagrams are half-ladder diagrams with fixed endpoints, which we choose to be the legs $+$ and $-$ associated to the loop momentum.}
\label{fig:half-ladder-1loop}
\end{figure}

\vspace{.3cm}
\paragraph{Trivalent diagrams and colour-kinematics duality.} We can now proceed as at tree level to translate the CHY-type expression for the loop integrand into trivalent diagrams. The analogues of \eqref{equ:Icoltreecf} and \eqref{equ:Ikintreept} are 
\begin{align}
\cI_{\text{SU}(N_c)}^{(1)} \; = \sum_{\rho\in S_{n}} 
\frac{ c^{(1)}\big(+,\rho(1,\cdots,n),-\big) 
}{\sigma_{+\rho(1)}\,\sigma_{\rho(1)\rho(2)} \cdots \sigma_{\rho(n)-} \,\sigma_{-+}} \,,
\label{equ:Icol1loopcf}
\end{align}
with
\begin{align}
c^{(1)}(\cdots) = \delta^{a_+a_-}\, c(\cdots) \,,
\end{align}
and
\begin{align}
E_a^{(1)}=0\quad \Rightarrow \quad \cI_{\text{kin}}^{(1)}\;=\sum_{\rho\in S_{n}} 
\frac{ N^{(1)}\big(+,\rho(1,\cdots,n),-\big) 
}{\sigma_{+\rho(1)}\,\sigma_{\rho(1)\rho(2)} \cdots \sigma_{\rho(n)-} \,\sigma_{-+}} \,,
\label{equ:Ikin1looppt}
\end{align}
with
\begin{align}
N^{(1)}(\cdots) = \sum_r N(\cdots) \,,
\end{align}
where the sum over states is again performed by the state projector \eqref{equ:1loopcompleteness}.
Our choice of master diagrams is pictured in \cref{fig:half-ladder-1loop}. From the master diagrams, all other trivalent diagrams can be obtained via the Jacobi relations.

In terms of trivalent diagrams, we have
\begin{equation}
\label{equ:cubic1loop}
 \mathcal{A}^{(1)}_{\text{YM}} = \int \frac{\d^D\ell}{\ell^2} \sum_{\ra\in \Gamma^{(1)}_{n+2}} \frac{N^{(1)}_\ra\, c^{(1)}_\ra}{D_\ra}\,,
 \qquad \qquad
  \mathcal{A}^{(1)}_{\text{grav}} = \int \frac{\d^D\ell}{\ell^2} \sum_{\ra\in \Gamma^{(1)}_{n+2}} \frac{N^{(1)}_\ra\, \tilde N^{(1)}_\ra}{D_\ra}\,.
\end{equation}
To interpret these formulae, let us recall that the loop-integrand representation directly obtained from the loop-level scattering equations has linear-type propagators \cite{Geyer:2015bja}.\footnote{See \cite{Gomez:2017lhy,*Gomez:2017cpe,*Ahmadiniaz:2018nvr} for an alternative approach.} Such a representation is obtainable from a Feynman-type representation with only quadratic propagators via partial fractions and shifts in the loop momentum \cite{Geyer:2015bja}. The partial fractions can be implemented via a residue argument \cite{Baadsgaard:2015twa}. As an example, an $n$-gon diagram with standard propagator structure is decomposed into $n$ tree-level-like diagrams with $n+2$ legs, as in the figure below. 
\vspace{.2cm}
\begin{center}
\begin{tikzpicture}[scale=0.38]
 \draw (-0.5,0.866) -- (0,0) -- (1,0) -- (2,0) -- (3,1.732) -- (2,3.464) -- (1,3.464);
 %-- (-0.5,2.598);
 \draw[dotted, thick] (-0.5,0.866) -- (-1,1.732) -- (0,3.464) -- (1,3.464);
 \draw (0,0) -- (-1,-1.732);
 \draw[dotted, thick] (-1,1.732) -- (-3,1.732);
 \draw[dotted, thick] (0,3.464) -- (-1, 5.196);
 \draw (3,1.732) -- (5,1.732);
 \draw (2,0) -- (3,-1.732);
 \draw (2,3.464) -- (3, 5.196);
 \node at (-1.4,-2.3) {\scalebox{0.7}{$i$}};
 \node at (3.5,-2.3) {\scalebox{0.7}{$i-1$}};
 
 \node[scale=0.9] at (8,1.4) {\scalebox{1}{$\displaystyle{ \;\sim} \quad\frac1{ { \ell^2}}\sum_i\,$\hspace{-.4cm}}};
 \draw (13.5,0.866) -- (14,0) -- (14.75,0);
 \draw (15.25,0) -- (16,0) -- (17,1.732) -- (16,3.464) -- (15,3.464);
 %-- (-0.5,2.598);
 \draw[dotted, thick] (13.5,0.866) -- (13,1.732) -- (14,3.464) -- (15,3.464);
 \draw (14,0) -- (13,-1.732);
 \draw[dotted, thick] (13,1.732) -- (11,1.732);
 \draw[dotted, thick] (14,3.464) -- (13, 5.196);
 \draw (17,1.732) -- (19,1.732);
 \draw (16,0) -- (17,-1.732);
 \draw (16,3.464) -- (17, 5.196);
 \draw[thick] (14.75,-0.7) -- (14.75,0.7);
 \draw[thick] (15.25,-0.7) -- (15.25,0.7);
 \node at (14.5,-1.3) {\scalebox{0.7}{$+$}};
 \node at (15.5,-1.3) {\scalebox{0.7}{$-$}};
 \node at (12.7,-2.3) {\scalebox{0.7}{$i$}};
 \node at (17.5,-2.3) {\scalebox{0.7}{$i-1$}};
 
 \node[scale=0.9] at (21.8,1.4) {\scalebox{1}{\,$\displaystyle =\quad \frac1{ \ell^2}\sum_i$\hspace{-.4cm}}};
 \draw (26,1) -- (28.25,1);
 \draw (27.5,1) -- (27.5,2.5);
 \draw[dotted,thick] (28.25,1) -- (31.25,1);
 \draw (31.25,1) -- (36.5,1);
 \draw[dotted,thick] (29,1) -- (29,2.5);
 \draw[dotted,thick] (30.5,1) -- (30.5,2.5);
 \draw (32,1) -- (32,2.5);
 \draw (33.5,1) -- (33.5,2.5);
 \draw (35,1) -- (35,2.5);
 \node at (25.4, 1) {\scalebox{0.7}{$+$}};
 \node at (37,1) {\scalebox{0.7}{$-$}};
 \node at (27.5,3) {\scalebox{0.7}{$i$}};
 \node at (35,3) {\scalebox{0.7}{$i-1$}};

 %\node at (40.2,1.4) {\scalebox{1}{$\in\;\displaystyle{\blue \mathfrak{I}_\ell}\,$}};
\end{tikzpicture}
\end{center}

\noindent The single quadratic propagator at one loop is explicit in \eqref{equ:cubic1loop}. For instance, for the diagram in fig.~\ref{fig:half-ladder-1loop}, the propagator set $1/D_\ra$ is
\begin{equation}
\label{propmap1loop}
\frac1{\prod_{i=1}^{n-1}(L+ K_i)^2} \quad \mapsto \quad \frac1{\prod_{i=1}^{n-1}\left(2\ell\cdot K_i+K_i^2\right)}\,, \qquad \text{with} \quad K_i=\sum_{j=1}^i k_{\rho(j)}\,.
\end{equation}
We use $\Gamma^{(1)}_{n+2}$ to denote the set of trivalent diagrams with $n+2$ legs, excluding only diagrams that contain a tadpole or an external-leg bubble after gluing legs $+$ and $-$. This exclusion is consistent with the result of the moduli space (CHY-type) integral, which is manifestly finite.
The naively singular contributions from the forward limit either cancel out or are finite but vanish upon loop integration in dimensional regularisation.

%%%%%%%%%%%%%%%%%%%%%%%%%%%%%%%%%%%%%%%%%%%%%%

\vspace{.3cm}
\subsection{Previous work at two loops}  \label{sec:review2loops}

\vspace{.3cm}
Similarly to the one-loop story, previous work on CHY-type formulae at two loops was based on ambitwistor strings, and applies to the loop integrand. Four-point formulae for maximal super-Yang-Mills theory and supergravity were presented in \cite{Geyer:2016wjx}. These were based on a heuristic derivation from ambitwistor strings at genus two, originally studied in \cite{Adamo:2015hoa}. A rigorous derivation via a residue argument in moduli space was presented more recently by two of the present authors \cite{Geyer:2018xwu}. This work also provided formulae at any multiplicity, although the `stringy' sum over spin structures was not fully simplified beyond four points. For other work at two loops, see also \cite{Feng:2016nrf,Gomez:2016cqb}.

The formulae for the loop integrands are similar to those for tree-level amplitudes, but now two nodes (four extra punctures) arise on the Riemann sphere, representing loop momenta insertions $\pm\ell_1$ and $\pm\ell_2$.  \\
\vspace{-.3cm}
\begin{figure}[h]
\begin{center}
\includegraphics[scale=1.5]{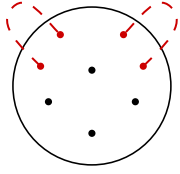} \\
\vspace{-.6cm}
\end{center}
\caption{Sphere with two nodes.}
\label{binodalsphere}
\end{figure}
\vspace{.1cm}\\
Let us label the punctures on the bi-nodal sphere by $\{\sigma_i,\sigma_{1^\pm},\sigma_{2^\pm}\}$. Remarkably, there are two possible sets of two-loop scattering equations \cite{Geyer:2016wjx}, defined by 
\begin{equation}
P^2(\sigma)-\ell_1^2\,\omega_1(\sigma)^2-\ell_2^2\,\omega_2(\sigma)^2+ \alpha\, (\ell_1^2+\ell_2^2)\,\omega_1(\sigma)\,\omega_2(\sigma)=0 \,, \qquad \boxed{\alpha=\pm 1}\,. 
\label{eq:P2_2loop}
\end{equation}
Here,
\begin{equation}
P^\mu(\sigma)=\ell_1^\mu\,\omega_1(\sigma)+\ell_2^\mu\,\omega_2(\sigma)+\sum_{i=1}^n \frac{k_i^\mu}{\sigma-\sigma_i}\,,
\quad \textrm{and} \quad
\omega_I(\sigma) = \frac{\sigma_{I^+I^-}}{(\sigma-\sigma_{I+})(\sigma-\sigma_{I^-})}\,.
\label{eq:P_2loop}
\end{equation}
Similarly to one loop, the $n$-point two-loop scattering equations can be written more clearly by considering null versions of the loop momenta: for $I=1,2$,
\begin{align}
\label{equ:2loopL}
L_I \qquad & \textrm{such that} \qquad L_I^2=0\,, \quad L_I\cdot k_i \,\mapsto\, \ell_I\cdot k_i \,, \quad L_I\cdot \epsilon_i \,\mapsto\, \ell_I\cdot \epsilon_i\,, \\
& \textrm{but also} \qquad \; 2L_1\cdot L_2 \,\mapsto\, \alpha\left(\ell_1+\alpha\,\ell_2\right)^2 \,. \nonumber
\end{align}
That is, the $n+4$ null momenta are $\{k_i,L_1,-L_1,L_2,-L_2\}$.

Therefore, the two-loop scattering equations are given by
\begin{equation}
E_a^{(2,+)}=0 \qquad \textrm{or} \qquad E_a^{(2,-)}=0\,,
\label{eq:SE2loopEpm}
\end{equation}
corresponding to $\alpha=1$ and $\alpha=-1$, respectively. Explicitly,
\begin{subequations} \label{equ:2loopSEell}
\begin{align}
E_i^{(2,\alpha)}&=k_i\cdot\ell_1\left(  \frac{1}{\sigma_i-\sigma_{1^+}} - \frac{1}{\sigma_i-\sigma_{1^-}} \right)
 +k_i\cdot\ell_2\left(  \frac{1}{\sigma_i-\sigma_{2^+}} - \frac{1}{\sigma_i-\sigma_{2^-}} \right)+\sum_{j\neq i} \frac{k_i\cdot k_j}{\sigma_i-\sigma_j} \,,\\
\pm E_{1^\pm}^{(2,\alpha)}&=\frac{\alpha}{2}\left(\ell_1+\alpha\, \ell_2\right)^2 \left(  \frac{1}{\sigma_{1^\pm}-\sigma_{2^+}} - \frac{1}{\sigma_{1^\pm}-\sigma_{2^-}} \right)+\sum_{j} \frac{\ell_1\cdot k_j}{\sigma_{1^\pm}-\sigma_j}\,,\\
\pm E_{2^\pm}^{(2,\alpha)}&=\frac{\alpha}{2}\left(\ell_1+\alpha\, \ell_2\right)^2 \left(  \frac{1}{\sigma_{2^\pm}-\sigma_{1^+}} - \frac{1}{\sigma_{2^\pm}-\sigma_{1^-}} \right)+\sum_{j} \frac{\ell_2\cdot k_j}{\sigma_{2^\pm}-\sigma_j}\,.
\end{align}
\end{subequations}

The crucial features appearing first at two loops are the quadratic factor $\left(\ell_1+\alpha\ell_2\right)^2$ and the two choices $\alpha=\pm1$, which are associated with aligned / anti-aligned loop momenta in diagrams.  The necessity of the quadratic factor was first noticed in \cite{Geyer:2016wjx} based on loop-integrand factorisation requirements, which coincide with those of `Q-cuts' \cite{Baadsgaard:2015twa}. For instance, one way to decompose a planar double-box diagram with standard propagator structure is represented below, in terms of three tree-level-like diagrams with $n+4$ external legs.
  \vspace{-.3cm}
\begin{center}
\begin{figure}[h]
\includegraphics[width=15cm]{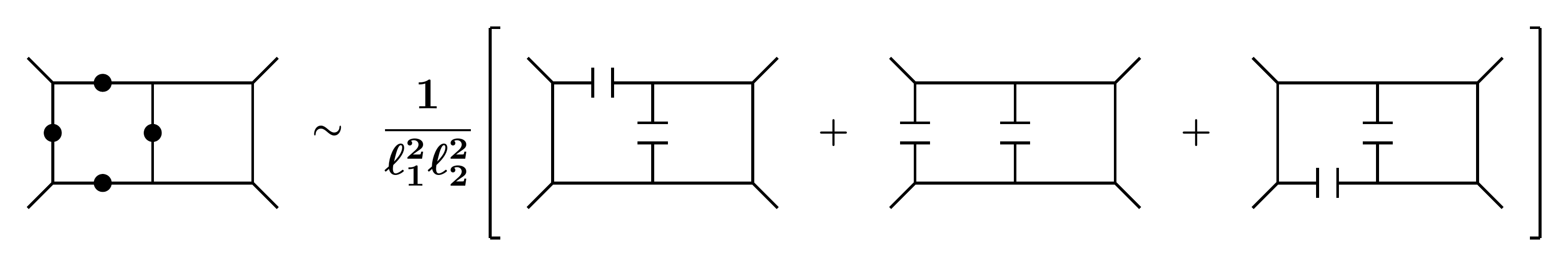} \\
  \vspace{-.5cm}
\caption{One of the ways of decomposing a planar double-box diagram in terms of tree-level diagrams with $n+4$ external legs. The dots on the left part denote three ways of placing one loop momentum (say $\ell_1$), which leads to the three diagrams on the right-hand side, whereas the dot in the middle denotes the other loop momentum.}
\label{2loopexample}
\end{figure}
\end{center}
  \vspace{-.6cm}
The second of these diagrams has the propagator structure
   {$$\displaystyle \frac1{{ (2\,\ell_1\cdot k_2)\,(-2\,\ell_1\cdot k_1)\,}{ (\ell_1+\ell_2+k_2)^2\,(\ell_1+\ell_2+k_2+k_3)^2\,(\ell_1+\ell_2-k_1)^2}}$$} 
if we choose the loop assignment as in the figure below. \\
  \vspace{-.6cm}
\begin{center}
\begin{figure}[h]
\hspace{6cm} \includegraphics[width=3cm]{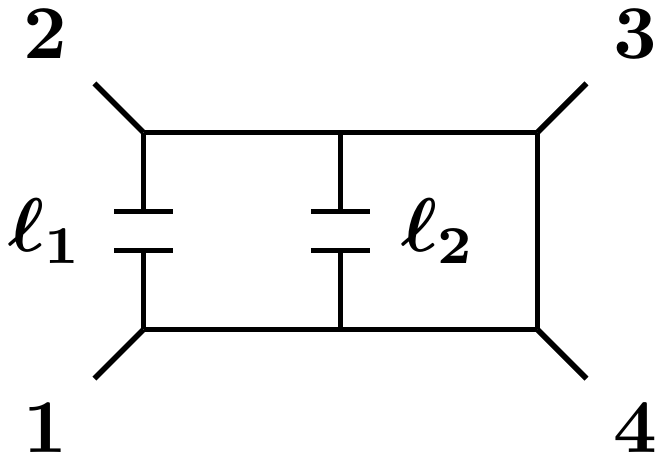} \\
\end{figure} 
\end{center}
  \vspace{-1.3cm}
As at one loop, such a representation can be obtained via partial fraction and shifts in the loop momenta. Analogously to \eqref{propmap1loop}, we have
\begin{equation}
\frac1{(L_1+\alpha\,L_2+K)^2}  \;\mapsto\;  \frac1{(\ell_1+\alpha\,\ell_2+K)^2} \,, \qquad
\text{for} \quad \alpha=\pm1\, .
\label{propmap2loop}
\end{equation}
That is, the propagators involving both loop momenta should be quadratic, unlike the ones involving only one of the loop momenta. Therefore, parallel / anti-parallel loop momenta are associated with $\alpha=1$ / $\alpha=-1$. In the next section, we will discuss the consequences of this fact. More examples of writing standard two-loop diagrams  in terms of the representation used here were given in \cite{Geyer:2016wjx}.

The derivation of expressions for supersymmetric theories from genus-two ambitwistor strings  illuminates the new two-loop features \cite{Geyer:2018xwu}. The inclusion of the quadratic factor is required for the residue argument, in order to fully localise the genus-two moduli space integration on the bi-nodal sphere of fig.~\ref{binodalsphere}. Regarding the two sets of scattering equations, only one is needed to describe two-loop amplitudes in the presence of supersymmetry (maximal in the case of gauge theory, and at least half-maximal in the case of gravity).

Let us conclude this section by commenting on singular solutions. Ref.~\cite{Geyer:2016wjx} discussed in detail the counting of regular and singular solutions to these equations (same for $\alpha=1$ and $\alpha=-1$). We will not need these solutions to solve the moduli space integral, for the same reason as at one loop, but let us mention for completeness that there are:
\begin{itemize}
\item $(n+1)!-4n!+4(n-1)!+6(n-3)!$ regular solutions, i.e., with $\sigma_{1^+},\sigma_{1^-},\sigma_{2^+},\sigma_{2^-}$ all distinct;
\item $2n\big((n-1)!-2(n-2)!\big)$ singular solutions for which either $\sigma_{1^+}=\sigma_{1^-}$ or $\sigma_{2^+}=\sigma_{2^-}$;
\item $(n-2)^2(n-3)!$ singular solutions for which both $\sigma_{1^+}=\sigma_{1^-}$ and $\sigma_{2^+}=\sigma_{2^-}$.
\end{itemize}
The total number of solutions is smaller than $(n+1)!$, which would be the naive expectation from the double-forward limit. This is analogous to the one-loop case. Also analogously to one loop, singular solutions give no contribution to the loop integrand in supersymmetric theories.

%%%%%%%%%%%%%%%%%%%%%%%%%%%%%%%%%%%%%%%%%%%%%%
%%%%%%%%%%%%%%%%%%%%%%%%%%%%%%%%%%%%%%%%%%%%%%

\section{New two-loop formulae}\label{sec:proposal} 

This section contains our main proposal. We present all-multiplicity formulae for the two-loop integrands of Yang-Mills theory and gravity, with or without supersymmetry, and also of other theories with an analogous CHY-type structure. While we use the tree-level and one-loop results as guidance, new important features appear at two loops, such that the double-forward limit is not straightforward for generic theories. In section~\ref{sec:NSsector}, we will compare in detail the supersymmetric and the non-supersymmetric cases.

\subsection{Naive guess}\label{sec:naiveprop}

We start by presenting a naive guess for how to write a two-loop integrand in terms of the scattering equations. Let us recall from section~\ref{sec:review2loops} that, at two loops, we have $n+4$ punctures labelled by $a=\{i, I^\pm\}$, as in fig.~\ref{binodalsphere}: $n$ for external particles, and two pairs for the loop momenta. There are two sets of scattering equations, $E_a^{(2,\alpha)}=0$, with $\alpha=\pm1$. In a democratic spirit, we consider contributions from both sets on equal footing. 
The naive guess is that the two-loop amplitudes take the generic form\footnote{As we shall discuss further, this is consistent with the supersymmetric case \cite{Geyer:2016wjx,Geyer:2018xwu}, for which the two contributions in \eqref{equ:2loopgeneric} differ only by $\ell_2\to-\ell_2$.}
\begin{equation}
\label{equ:2loopgeneric}
 \mathcal{A}^{(2)}\stackrel{?}{=}\int \frac{\d^D\ell_1\,\d^D\ell_2}{\ell^2_1\,\ell^2_2} \int_{\mathfrak{M}_{0,n+4}}\! \frac1{2}\left[ \d\mu_{n+4}^{(2,+)}\;\cI^{(2,+)} + d\mu_{n+4}^{(2,-)}\;\cI^{(2,-)} \right] \,,
\end{equation}
with the measures
\begin{equation}
\label{equ:2loopSEmeasure}
\d\mu_{n+4}^{(2,\alpha)}\equiv \frac{\d^{n+4}\sigma}{\vol\SL(2,\C)} \ \prod_a{}' \, \delta(E_a^{(2,\alpha)})\,.
\end{equation}

In the following, we will describe how to construct the objects $\cI^{(2,\alpha)}$. Our starting point is the basic observation that, similarly to one loop, the $n$-point two-loop scattering equations can be defined from their tree-level $(n+4)$-point counterparts.  This is achieved by considering null versions of the loop momenta, as discussed in \eqref{equ:2loopL}.
This basic observation leads to a natural two-loop extension of the CHY Pfaffian, which would naively seem obstructed by the appearance of $\left(\ell_1\pm\alpha\ell_2\right)^2$ in the two-loop scattering equations. In analogy to the one-loop case, we can write
\begin{align}
\label{equ:Ikin2looppf}
\cI_{\text{kin}}^{(2,\alpha)} (\{\epsilon_i,k_i,\sigma_i\},\{\ell_I,\sigma_{I^\pm}\}) &= \sum_{r_1,r_2}\, \cI_{\text{kin}}^{(0)} (\{\epsilon_i,k_i,\sigma_i\},\{\epsilon_{I^\pm}^{r_I},\pm L_I,\sigma_{I^\pm}\}) \nonumber  \\
& = \sum_{r_1,r_2}\, \pf'(M) %= \Delta_1^{\mu\nu}\Delta_2^{\alpha\beta}\, \pf'(M)_{\mu\nu,\alpha\beta}\,.
= \Delta_{1\,\mu_1\nu_1}\Delta_{2\,\mu_2\nu_2}\, \pf'(M)^{\mu_1\nu_1,\mu_2\nu_2}\,.
\end{align}
The sum over the states running in the loop $I$ is defined by the state projector,
\begin{equation}
\label{equ:2loopcompleteness}
\sum_{r_I}\, \epsilon_{I^+\,\mu}^{\,r_I}\, \epsilon_{I^-\,\nu}^{\,r_I} = \eta_{\mu\nu} - \frac{L_{I\mu} q_\nu+L_{I\nu} q_\mu}{L_I\cdot q} \equiv \Delta_{I\mu\nu} \,,
\end{equation}
where $q$ is a null reference vector. For complete clarity, the object $\pf'(M)^{\mu_1\nu_1,\mu_2\nu_2}$ is defined precisely as at tree level, from
\begin{equation}
 M=\begin{pmatrix}A &-C^T\\C&B\end{pmatrix}\,,
\end{equation}
which is now a $2(n+4)\times2(n+4)$ matrix. The reduced Pfaffian is still
\begin{equation}
 \pf{}'\left(M\right)=\frac{(-1)^{ \hat{a}+ \hat{b}}}{\sigma_{\hat{a}\hat{b}}}\pf\left(M^{\hat{a}\hat{b}}_{\hat{a}\hat{b}}\right)\,,
\end{equation}
where $M^{\hat{a}\hat{b}}_{\hat{a}\hat{b}}$ denotes the matrix $M$ with both rows and columns $\hat{a}$ and $\hat{b}$ removed (for $\hat{a}<\hat{b}\leq n+4$). Moreover,
\begin{equation}
\pf'(M)_{\mu_1\nu_1,\mu_2\nu_2}= \partial_{\epsilon_{1^+}^{\mu_1}} \partial_{\epsilon_{1^-}^{\nu_1}} \partial_{\epsilon_{2^+}^{\mu_2}} \partial_{\epsilon_{2^-}^{\nu_2}} \pf{}'\left(M\right) \,.
\end{equation}
Explicitly, the components of $M$ are: for the $A$ submatrix,
\begin{subequations}
\begin{align}
 &A_{I^+I^-}=0 && A_{1^\pm2^+}=\pm\frac{\alpha\,\left(\ell_1+\alpha\,\ell_2\right)^2}{2\,\sigma_{1^\pm2^+}}&&  A_{1^\pm2^-}=\mp\frac{\alpha\,\left(\ell_1+\alpha\,\ell_2\right)^2}{2\,\sigma_{1^\pm2^-}}\\
 & && A_{I^\pm \,j}=\pm\frac{\ell_I\cdot k_j}{\sigma_{I^\pm j}} && A_{ij}=\frac{k_i\cdot k_j}{\sigma_{ij}}\,,
 \end{align}
\end{subequations}
where the factors $\frac1{2}\left(\ell_1+\ell_2\right)^2$ appear from $L_1\cdot L_2$; for the $B$ submatrix,
\begin{subequations}
\begin{align}
 &B^{\mu_I\nu_I}_{I^+I^-}=\frac{\eta^{\mu_I\nu_I}}{\sigma_{I^+I^-}} && B^{\mu_I\mu_J}_{I^+J^+}=\frac{\eta^{\mu_I\mu_J}}{\sigma_{I^+J^+}} && B^{\mu_I\nu_J}_{I^+J^-}=\frac{\eta^{\mu_I\nu_J}}{\sigma_{I^+J^-}}\\
 & B^{\mu_I }_{I^+j}=\frac{\epsilon_j^{\mu_I}}{\sigma_{I^+j}} &&  B^{\nu_I }_{I^-j}=\frac{\epsilon_j^{\nu_I}}{\sigma_{I^-j}} && B_{ij}=\frac{\epsilon_i\cdot\epsilon_j}{\sigma_{ij}}\,;
 \end{align}
\end{subequations}
and for the $C$ submatrix,
\begin{subequations}
\begin{align}
 &C_{I^+I^-}=0 && C^{\mu_I}_{I^+J^\pm} =\pm\frac{L_J^{\mu_I}}{\sigma_{I^+J^\pm}} && C^{\mu_I}_{I^+j} =\frac{k_j^{\mu_I}}{\sigma_{I^+j}}\\
 &&& C^{\nu_I}_{I^-J^\pm} =\pm\frac{L_J^{\nu_I}}{\sigma_{I^-J^\pm}} && C^{\nu_I}_{I^-j} =\frac{k_j^{\nu_I}}{\sigma_{I^-j}}\\
 & C_{ iI^\pm}=\pm\frac{\epsilon_i\cdot \ell_I }{\sigma_{iI^\pm}}&& C_{ij}=\frac{\epsilon_i\cdot k_j}{\sigma_{ij}} &&C_{aa}=-\sum_{b\neq a}C_{ab}\,.
\end{align}
\end{subequations}
Just as at tree-level and at one-loop, the full Pfaffian $\sum_{r_1,r_2}\pf\left(M\right)$ vanishes on the support of the scattering equations, with the kernel spanned by the vectors $(1,\dots,1,0,\dots,0)$ and $(\sigma_{1^+},\dots,\sigma_n,0,\dots,0)$, so the reduced Pfaffian is well-defined and invariant under the choice of removed rows and columns.

The kinematic object defined above, starting in \eqref{equ:Ikin2looppf}, is an important piece in our construction. 
It is manifestly invariant for gauge transformations of the external polarisations, if evaluated on solutions to the two-loop scattering equations that are regular (i.e., where $\sigma_{1^+},\sigma_{1^-},\sigma_{2^+},\sigma_{2^-}$ are all distinct), for the same reason that the analogous statement holds at tree level. 
Likewise, the choice of the reference vector $q$ in \eqref{equ:2loopcompleteness} is arbitrary.
In fact, we can use an effective substitution rule that avoids its use altogether: for $V,W\in\{k_i,\epsilon_i\}$,
\begin{align}
\label{equ:2loopcompletenessrule}
&\Delta_{I\mu\nu} \,V^\mu W^\nu\,,\;
\Delta_{I\mu}{}^\alpha\Delta_{J\alpha\nu} \,V^\mu W^\nu \quad
\leadsto \quad V\cdot W \,, \nonumber\\
&\Delta_{I\mu\nu} \,L_J^\mu V^\nu\,,\;
\Delta_{I\mu}{}^\alpha\Delta_{J\alpha\nu} \,L_J^\mu V^\nu \quad
\leadsto\quad V\cdot (L_J-L_I) \,\mapsto\, V\cdot (\ell_J-\ell_I) \,,  \\
&\Delta_{1\mu\nu} \,L_2^\mu L_2^\nu\,,\;
 \Delta_{2\mu\nu} \,L_1^\mu L_1^\nu \,,\;
 - \Delta_{1\mu}{}^\alpha\Delta_{2\alpha\nu} \,L_2^\mu L_1^\nu \quad
 \leadsto\quad -2\,L_1\cdot L_2 \,\mapsto\, -\alpha(\ell_1+\alpha\,\ell_2)^2\,, \nonumber
\end{align}
while $\Delta_{I\mu}{}^\mu =\Delta_{1\mu\nu}\,\Delta_{2}^{\mu\nu} =D-2$\,. We made explicit numerical checks of the substitution rule by evaluating the kinematic object \eqref{equ:Ikin2looppf} on solutions to the scattering equations.

Having defined the kinematic object \eqref{equ:Ikin2looppf}, we can consider also its colour counterpart. This was first discussed in \cite{Geyer:2018xwu} in the context of super-Yang-Mills theory. Only the kinematic object depends on the degree of supersymmetry, not the colour object. The idea is to simply glue the colour indices for each loop, starting from the tree-level $(n+4)$-point object,
\begin{equation}
\label{equ:Icol2loopfl}
\cI_{\text{SU}(N_c)}^{(2)} (\{a_i,\sigma_i\},\{\sigma_{I^\pm}\}) = \delta^{a_{1^+}a_{1^-}}\,\delta^{a_{2^+}a_{2^-}}\, \cI_{\text{SU}(N_c)}^{(0)} (\{a_i,\sigma_i\},\{a_{I^\pm},\sigma_{I^\pm}\}) \,,
\end{equation}
which is independent of the choice $\alpha=\pm1$. This expression contains the full colour dependence, including non-planar contributions. If one wants to decompose the colour dependence in a $1/N_c$ expansion, this is as usual accomplished with the completeness relation for the Lie algebra of SU($N_c$),
\begin{equation}
 (T^a)_{i_1}^{\;\;j_1}\,(T^a)_{i_2}^{\;\;j_2}= \delta_{i_1}^{j_2}\delta_{i_2}^{j_1}-\frac{1}{N_c}\,\delta_{i_1}^{j_1}\delta_{i_2}^{j_2} \,.
\end{equation}
The expression \eqref{equ:Icol2loopfl} does not yet coincide with the one presented in \cite{Geyer:2018xwu}, denoted as $ \cI_{\text{SU}(N_c)}^{(2,+)}$ below.

We now proceed to the naive-guess formulae for gauge theory and gravity, based on  \eqref{equ:2loopgeneric}. They exhibit a double-copy structure just like at tree level and one loop, but have important novel features. In fact, we propose two representations of the formulas. The two representations are only meant to agree when we consider the sum over the contributions from both sets of scattering equations in \eqref{equ:2loopgeneric}.
 The first representation is
\begin{align}
\boxed{
 \qquad \cI_{\text{YM}}^{(2,\alpha)}=\xi^{(\alpha)}\,\cI_{\text{kin}}^{(2,\alpha)}\,\, \cI_{\text{SU}(N_c)}^{(2)}\,,\qquad\qquad \cI_{\text{grav}}^{(2,\alpha)}=\xi^{(\alpha)}\,\cI_{\text{kin}}^{(2,\alpha)}\,\,\widetilde\cI_{\text{kin}}^{(2,\alpha)}\,,\qquad}
 \label{eq:CHY_YM_grav_2loop_cr}
\end{align}
with the cross-ratios
\begin{equation}
\xi^{(+)}=\frac{\sigma_{1^+2^-}\,\sigma_{2^+1^-}}{\sigma_{1^+1^-}\,\sigma_{2^+2^-}} \,,
\qquad
\xi^{(-)}=\frac{\sigma_{1^+2^+}\,\sigma_{2^-1^-}}{\sigma_{1^+1^-}\,\sigma_{2^-2^+}} \,,
\end{equation}
such that
\begin{equation}
\xi^{(+)}+\xi^{(-)}=1\,.
\end{equation}
Similarly to tree level and one loop, $\widetilde\cI_{\text{kin}}^{(2,\alpha)}$ coincides with $\cI_{\text{kin}}^{(2,\alpha)}$ after substituting the set of external polarisations, $\epsilon_i\to\tilde\epsilon_i$.

The second representation, which we denote with a prime, is
\begin{align}
\boxed{
 \qquad \cI'{}_{\text{YM}}^{(2,\alpha)}=\cI_{\text{kin}}^{(2,\alpha)}\,\, \cI_{\text{SU}(N_c)}^{(2,\slashed{\alpha})}\,,\qquad\qquad \cI'{}_{\text{grav}}^{(2,\alpha)}=\cI_{\text{kin}}^{(2,\alpha)}\,\,\widetilde\cI_{\text{kin}}^{(2,\slashed{\alpha})}\,.\qquad}
 \label{eq:CHY_YM_grav_2loop_slash}
\end{align}
We use the $\slashed{\alpha}$ symbol, instead of just $\alpha$, to denote the following. Consider the definition of $\cI_{\text{SU}(N_c)}^{(2)}$, which is actually independent of $\alpha$, written in terms of Lie algebra structure constants mirroring the tree-level case \eqref{equ:Icoltreecf},
\begin{equation}
\cI_{\text{SU}(N_c)}^{(2)} = \delta^{a_{1^+}a_{1^-}}\,\delta^{a_{2^+}a_{2^-}} \sum_{\gamma \in S_{n+2}} 
\frac{f^{a_{1^+}a_{\gamma(1)}b_1} \, f^{b_1a_{\gamma(2)}b_2} \cdots f^{b_{n-3}a_{\gamma(n+2)}a_{1^-}} }
{\sigma_{1^+\gamma(1)}\,\sigma_{\gamma(1)\gamma(2)} \cdots \sigma_{\gamma(n+2)1^-} \,\sigma_{1^-1^+}}\,,
\label{equ:Icol2loopcf}
\end{equation}
where $S_{n+2}$ is the set of permutations of $\{i,2^\pm\}$, i.e., $\{1^\pm\}$ are fixed. Then the `slashed' version is
\begin{equation}
\cI_{\text{SU}(N_c)}^{(2,\slashed{\alpha})} = \delta^{a_{1^+}a_{1^-}}\,\delta^{a_{2^+}a_{2^-}} \sum_{\gamma \in S^{({\alpha})}_{n+2}} 
\frac{f^{a_{1^+}a_{\gamma(1)}b_1} \, f^{b_1a_{\gamma(2)}b_2} \cdots f^{b_{n-3}a_{\gamma(n+2)}a_{1^-}} }
{\sigma_{1^+\gamma(1)}\,\sigma_{\gamma(1)\gamma(2)} \cdots \sigma_{\gamma(n+2)1^-} \,\sigma_{1^-1^+}}\,.
\label{equ:hatIcol2loopcf}
\end{equation}
The distinction is that we consider now a restricted set of permutation $S^{({\alpha})}_{n+2}$, with half the elements of $S_{n+2}$, such that any $\gamma \in S^{(\pm)}_{n+2}$ is of the type $\{\cdots, 2^\pm, \cdots, 2^\mp, \cdots\}$. Clearly,
\begin{equation}
\cI_{\text{SU}(N_c)}^{(2)}=\sum_{\alpha=\pm} \cI_{\text{SU}(N_c)}^{(2,\slashed{\alpha})} \,.
\end{equation}
The `slash' prescription may also be applied to a kinematic object. Given the double-forward-limit interpretation of $\cI_{\text{kin}}^{(2)}$, we have
\begin{align}
E_a^{(2,\alpha)}=0\quad \Rightarrow \quad \cI_{\text{kin}}^{(2,\alpha)}\;=\sum_{\gamma\in S_{n+2}} 
\frac{ N^{(2)}\big(1^+,\gamma(1,\cdots,n,2^+,2^-),1^-\big) 
}{\sigma_{1^+\gamma(1)}\,\sigma_{\gamma(1)\gamma(2)} \cdots \sigma_{\gamma(n+2)1^-} \,\sigma_{1^-1^+}}\,,
\label{equ:Ikin2loopptA}
\end{align}
with
\begin{align}
N^{(2,\alpha)}(\cdots) = \sum_{r_1,r_2} N(\cdots) \,,
\label{equ:num2loops}
\end{align}
where the $N$ are tree-level numerators, as in \eqref{equ:Ikintreept}, and the sum over states running in the loops is performed by the state projectors \eqref{equ:2loopcompleteness}. It is therefore natural to define
\begin{align}
\cI_{\text{kin}}^{(2,\slashed{\alpha})}\;=\sum_{\gamma\in S^{(\alpha)}_{n+2}} 
\frac{ N^{(2)}\big(1^+,\gamma(1,\cdots,n,2^+,2^-),1^-\big) 
}{\sigma_{1^+\gamma(1)}\,\sigma_{\gamma(1)\gamma(2)} \cdots \sigma_{\gamma(n+2)1^-} \,\sigma_{1^-1^+}}\,,
\label{equ:hatIkin2loopptA}
\end{align}
where we restrict the sum over permutations, as in \eqref{equ:hatIcol2loopcf}. With this object in hand, notice that we could have `slashed' $\cI_{\text{kin}}^{(2)}$ instead of $\cI_{\text{SU}(N_c)}^{(2)}$ in the gauge theory formula  of \eqref{eq:CHY_YM_grav_2loop_slash}.

In both representations of our formulae, we are splitting an expression into two parts, either by the cross-ratio as in \eqref{eq:CHY_YM_grav_2loop_cr} or by dropping half the terms in each part as in \eqref{eq:CHY_YM_grav_2loop_slash}. Each of the two parts is evaluated in \eqref{equ:2loopgeneric} with the corresponding set of scattering equations. The reason for this restriction is that these equations pick a relative orientation of the loop momenta, via the factor $(\ell_1+\alpha\ell_2)^2$. The scattering equations admit loop-integrand-level poles of the type
\begin{equation}
\frac1{(L_1\pm L_2+K)^2}
\quad \mapsto \quad
\frac1{\pm\alpha(\ell_1+\alpha\,\ell_2)^2+2(\ell_1\pm\ell_2)\cdot K+K^2}\,,
\end{equation}
where $K$ is a sum of external momenta. The suppression of factorisation-forbidden poles (such that $\ell_1\pm\ell_2$ is $\ell_1-\ell_2$ for $\alpha=+1$, and $\ell_1+\ell_2$ for $\alpha=-1$) is a special property of valid CHY-type integrands. Both versions of our formulae, \eqref{eq:CHY_YM_grav_2loop_cr} and \eqref{eq:CHY_YM_grav_2loop_slash}, satisfy the requirement of suppressing these poles, and allowing the correct poles,
\begin{equation}
\frac1{(L_1+\alpha L_2+K)^2}
\quad \mapsto \quad
\frac1{(\ell_1+\alpha\,\ell_2+K)^2} \,.
\end{equation}
It is not hard to see how they may satisfy this requirement.\footnote{It may be helpful to recall here the relation between worldsheet factorisation and kinematic factorisation. Let us consider the tree-level scattering equations. If some marked points merge as solutions to the scattering equations, $\sigma_{i\in A}\to \sigma_\ast$, then this implies that $(\sum_{i\in A}k_i)^2\to0$. If the moduli-space integration encounters a pole as the marked points merge, then the (integrated) expression will possess the associated kinematic pole.} For instance, the cross-ratio $\xi^{(+)}$ in \eqref{eq:CHY_YM_grav_2loop_cr} suppresses worldsheet degenerations where $\sigma_{1^+}=\sigma_{2^-}$ or $\sigma_{1^-}=\sigma_{2^+}$, since then $\xi^{(+)}=0$.
These degenerations would lead to factorisation-forbidden poles when associated to %the factors $(\ell_1+\ell_2)^2$ in 
the $E^{(2,+)}_a=0$ scattering equations. The `slash' prescription achieves the same goal, since the ordering of the loop punctures in the Parke-Taylor denominators, given by the set of permutations $S^{(+)}_{n+2}$ for $E^{(2,+)}_a=0$, does not lead to poles for $\sigma_{1^+}=\sigma_{2^-}$ or $\sigma_{1^-}=\sigma_{2^+}$.

Apart from Yang-Mills theory and gravity, it is also natural to consider other theories, e.g., those in \cite{Cachazo:2014xea}. The naive-guess formulae in those cases should also be clear. The double-forward limit of the loop momenta can be performed with either $\alpha=1$ or $\alpha=-1$. In the cross-ratio prescription, that is all that is needed. In the `slash' prescription, one first identifies a Parke-Taylor decomposition analogous to \eqref{equ:Icol2loopcf} and \eqref{equ:hatIkin2loopptA}, and then drops half the terms according to $\alpha$. For instance, for the bi-adjoint scalar theory we have simply
\begin{align}
\boxed{
 \qquad \cI_{\text{bi-adj}}^{(2,\alpha)}=\xi^{(\alpha)}\,\cI_{\text{SU}(N_c)}^{(2)}\,\, \cI_{\text{SU}(\tilde N_c)}^{(2)}\qquad}
 \label{eq:badj_2loop_cr}
\end{align}
for the cross-ratio prescription, and
\begin{align}
\boxed{
 \qquad \cI'{}_{\text{bi-adj}}^{(2,\alpha)}=\cI_{\text{SU}(N_c)}^{(2)}\,\, \cI_{\text{SU}(\tilde N_c)}^{(2,\slashed{\alpha})}\qquad}
 \label{eq:badj_2loop_slash}
\end{align}
for the `slash' prescription.

We verified the equivalence of the two prescriptions by numerically evaluating \eqref{equ:2loopgeneric} over the regular solutions to the scattering equations
 (the two sets) at four points in the case of the bi-adjoint scalar theory, i.e., using either \eqref{eq:badj_2loop_cr} or \eqref{eq:badj_2loop_slash}. We also verified the equivalence numerically at four points for super-Yang-Mills theory, where the object $\cI_{\text{kin}}^{(2,{\alpha})}$ is substituted by its supersymmetric counterpart -- more on this later.

While the two prescriptions have been verified in examples to be equivalent, there remains to be checked whether they lead to a valid two-loop integrand. Generically, this is not the case, and a correction to the naive guess \eqref{equ:2loopgeneric} is required, as we will discuss below.

\subsection{Trivalent diagrams and colour-kinematics duality}\label{sec:trivalent}

The naive guess \eqref{equ:2loopgeneric} leads to explicit expressions for the loop integrands. The idea is to use the same procedure as at tree level and one loop, reviewed in section~\ref{sec:review}. We recall that the representation of the loop integrands arising naturally from the scattering equations formalism contains linear-type propagators, not just quadratic ones. 

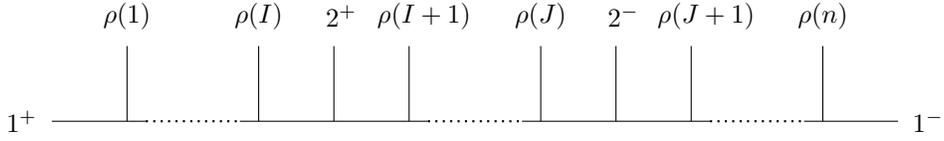
\begin{figure}[t]
\begin{center}
 \begin{tikzpicture}[scale=1]
 \draw (0,0) -- (1.25,0) ;
 \draw (1,0) -- (1,1) ;
 \draw[dotted, thick] (1.25,0) -- (2.5,0) ;
 \draw (2.75,0) -- (2.75,1) ; 
 \draw (3.75,0) -- (3.75,1) ; 
 \draw (4.75,0) -- (4.75,1) ; 
 \draw (2.5,0) -- (5,0);
 \draw[dotted, thick] (5,0) -- (6.25,0) ;
 \draw (6.25,0) -- (8.75,0) ;
 \draw (6.5,0) -- (6.5,1) ; *
 \draw (7.5,0) -- (7.5,1) ; *
 \draw (8.5,0) -- (8.5,1) ; *
 \draw[dotted, thick] (8.75,0) -- (10,0) ;
 \draw (10,0) -- (11.25,0) ;
 \draw (10.25,0) -- (10.25,1) ; *
 \node at (-0.4,0) {$1^+$} ;
 \node at (11.65,0) {$1^-$} ;
 \node at (1,1.4) {$\rho(1)$} ;
 \node at (2.75,1.4) {$\rho(I)$} ;
 \node at (3.85,1.4) {$2^+$} ;
 \node at (4.95,1.4) {$\rho(I+1)$} ;
 \node at (6.5,1.4) {$\rho(J)$} ;
 \node at (7.6,1.4) {$2^-$} ;
 \node at (8.7,1.4) {$\rho(J+1)$} ;
 \node at (10.25,1.4) {$\rho(n)$} ;
\end{tikzpicture}
\end{center}
\caption{The two-loop BCJ master diagrams are half-ladder diagrams with fixed endpoints, which we choose to be the legs $1^+$ and $1^-$ associated to the loop momentum 1. Half of the master diagrams have the relative ordering of legs $2^+$ and $2^-$ represented in this figure, and they arise from the first part of \eqref{equ:2loopgeneric}, associated to the scattering equations $E^{(2,+)}_a=0$. The second half of the master diagrams have the other relative ordering of legs $2^+$ and $2^-$, and are associated to the scattering equations $E^{(2,-)}_a=0$.}
\label{fig:half-ladder-2loop}
\end{figure}

In principle, the novel two-loop feature of the two sets of scattering equations does not complicate  the moduli space integrals. We expect, for our naive guess \eqref{equ:2loopgeneric}, that
\begin{equation}
\label{equ:cubic2loop}
 \mathcal{A}^{(2)}_{\text{YM}} \stackrel{?}{=} \int \frac{\d^D\ell_1\,\d^D\ell_2}{\ell^2_1\,\ell^2_2} \sum_{\ra\in \Gamma^{(2)}_{n+4}} \frac{N^{(2)}_\ra\, c^{(2)}_\ra}{D_\ra}\,,
 \qquad \qquad
  \mathcal{A}^{(2)}_{\text{grav}} \stackrel{?}{=} \int \frac{\d^D\ell_1\,\d^D\ell_2}{\ell^2_1\,\ell^2_2} \sum_{\ra\in \Gamma^{(2)}_{n+4}} \frac{N^{(2)}_\ra\, \tilde N^{(2)}_\ra}{D_\ra}\,.
\end{equation}
Here, $\Gamma^{(2)}_{n+4}$ denotes the set of tree-level trivalent diagrams with $n+4$ legs, excluding diagrams that contain one- or two-loop-type tadpoles or external-leg bubbles after gluing the legs corresponding to the loops ($1^+$ with $1^-$, and $2^+$ with $2^-$).\footnote{Notice that this includes also diagrams that possess an external-leg bubble when only one of the loop momenta is glued. The prescription for the exclusion of the bubble- and tadpole-diagrams is analogous to one loop, where they were shown to be associated to singular solutions, which at most give finite contributions to the loop integrand (and integrate to zero at one loop).} The set of trivalent diagrams is obtained from $(n+2)!$ master diagrams via Jacobi relations, analogously to tree level and one loop; see fig.~\ref{fig:half-ladder-2loop}. 
The effect of the splitting of \eqref{equ:2loopgeneric} into contributions from the two sets of scattering equations is merely to ensure the correct form of the quadratic propagators involving both loop momenta, that is, to enforce \eqref{propmap2loop}.
For instance, the diagrams pictured in fig.~\ref{fig:half-ladder-2loop} have propagator sets $1/D_\ra$ given by
\begin{align}
& \frac1{\left(\prod_{i=1}^{I}(L_1+ K_i)^2\right)
\left(\prod_{i=I}^{J}(L_1+L_2+ K_i)^2\right)
\left(\prod_{i=J}^{n{\phantom{I}}}(L_1+ K_i)^2 \right)} 
\nonumber \\ & \mapsto \quad 
\frac1{\left(\prod_{i=1}^{I}(2\ell_1\cdot K_i+K_i^2)\right)
\left(\prod_{i=I}^{J}(\ell_1+\ell_2+ K_i)^2\right)
\left(\prod_{i=J}^{n{\phantom{I}}}(2\ell_1\cdot K_i+K_i^2) \right)}
\,, \nonumber \\ 
& \text{with} \quad K_i=\sum_{j=1}^i k_{\rho(j)}\,.
\label{equ:example2loopprop}
\end{align}
The representation of the loop integrand is therefore not in a Feynman form. Instead, it contains linear-type propagators, as discussed in section~\ref{sec:review2loops}.
We recall that the colour factors and kinematic numerators of master diagrams are the ones appearing in \eqref{equ:hatIcol2loopcf} and \eqref{equ:hatIkin2loopptA}, respectively. 

As at tree level and one loop, the colour-kinematics duality holds by construction. It follows from tree level via the double-forward limit. This is in contrast with the original BCJ loop-level conjecture \cite{Bern:2010ue}, which was formulated for a representation with Feynman-type propagators, for which the implications of the well-established tree-level colour-kinematics duality are not fully understood. In the case of pure Yang-Mills theory at two loops, the original conjecture faces obstructions already at five points \cite{Bern:2015ooa,*Mogull:2015adi}, whereas the construction presented above is conjectured to be valid for any multiplicity.

\begin{figure}[t]
\includegraphics[width=15cm]{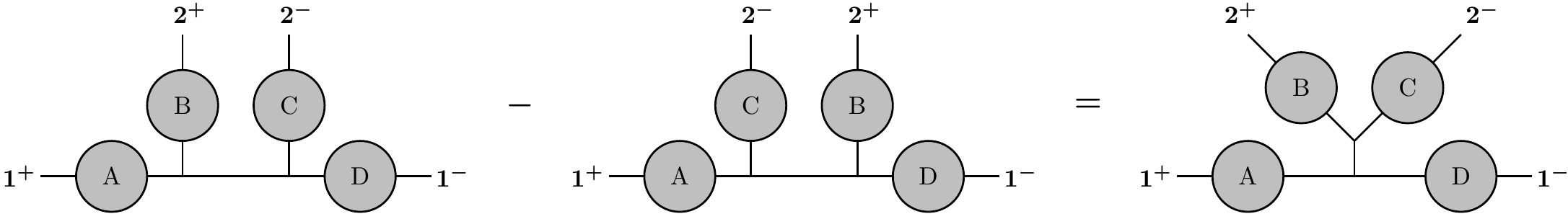} \\
  \vspace{-.5cm}
\caption{One class of Jacobi relations between trivalent diagrams at two loops. The `blobs' (A, B, C, D) are trivalent trees where we suppress the $n$ external legs. On the left-hand side, the diagrams contain a single quadratic propagator each, of the type $1/(\ell_1+\ell_2+K)^2$ in the first case and $1/(\ell_1-\ell_2+K)^2$ in the second case. On the right-hand side, the diagram has independent loops, and therefore possesses no propagator involving both loop momenta.}
\label{jacobi2loop}
\end{figure}

Before proceeding, let us relate the present discussion, which employs two sets of scattering equations, to previous work at two loops \cite{Geyer:2016wjx,Geyer:2018xwu}, which employed a single set of scattering equations (then chosen to be $\alpha=+1$). The distinction is that previous work considered only supersymmetric theories. Consider the class of Jacobi relations among tree-level-type diagrams represented in fig.~\ref{jacobi2loop}. It involves a diagram with parallel loop momenta, another with anti-parallel loop momenta, and on the right-hand side a diagram with independent loops. If the latter type of diagram contributes to the amplitude, then it is clear that we need both sets of scattering equations, since the kinematic numerators and colour factors of such diagrams are built from master numerators of both contributions in \eqref{equ:2loopgeneric}. It turns out that, in the supersymmetric case, diagrams with independent loop momenta do not contribute, as their kinematic numerators vanish. Indeed, the two contributions in \eqref{equ:2loopgeneric} are then equivalent, up to $\ell_2\to-\ell_2$, and one can use a single set of scattering equations, as in  \cite{Geyer:2016wjx,Geyer:2018xwu}. One important detail is that, in the case of supersymmetric gauge theory, this requires the use of the `slash' prescription of \eqref{eq:CHY_YM_grav_2loop_slash}, with the supersymmetric version of the kinematic object $\cI_{\text{susy-kin}}^{(2,\alpha)}$, and with the `slash' applied to the colour object $\cI_{\text{SU}(N_c)}^{(2,\alpha)}$. This prescription effectively eliminates  the would-be colour factors of half the diagrams, including that in the middle of fig.~\ref{jacobi2loop}, but preserves the kinematic numerators, such that the kinematic numerator of the diagram on the right-hand side still vanishes via the Jacobi relation, as it should. In the case of supergravity, the two prescriptions -- cross-ratio and `slash' -- are identical. As we will see in \cref{sec:cr}, the cross-ratio is equivalent to the `slash' of the supersymmetric kinematic object.  

\subsection{Failure of the naive guess}\label{sec:counting}

The naive guess \eqref{equ:cubic2loop} exhibits the double-forward limit and the colour-kinematics duality. It turns out, however, that it is incorrect. The reason is that we neglected to appropriately take into account the relation between the tree-like diagrams with $n+4$ legs and the standard diagrams with quadratic propagators that they are supposed to reproduce (up to loop integration).

\begin{figure}[t]
\includegraphics[width=15cm]{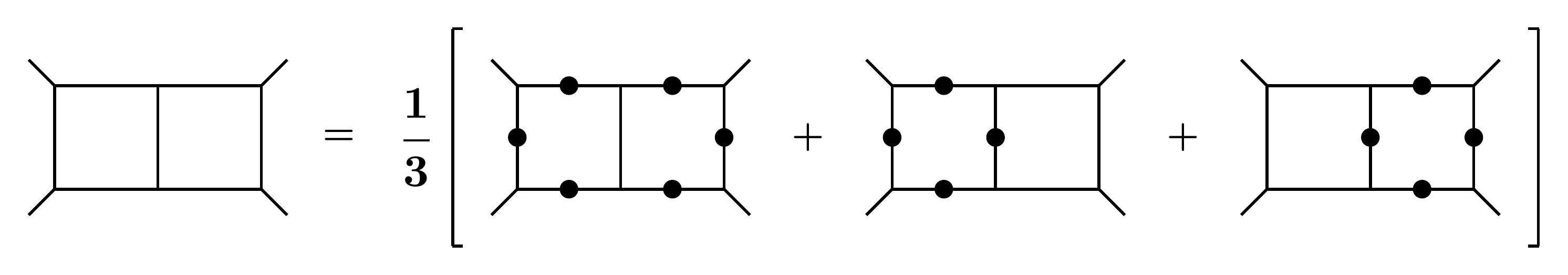} \\
  \vspace{-.5cm}
\caption{A double-box can be reconstituted in three distinct ways from $(n+4)$-pt tree diagrams. The second way was explicitly considered in section~\ref{sec:review2loops}. The first way corresponds to nine tree-like diagrams, whereas the second and third correspond to three.}
\label{2loopcounting1}
\end{figure}

\begin{figure}[t]
  \hspace{3cm}
\includegraphics[width=10cm]{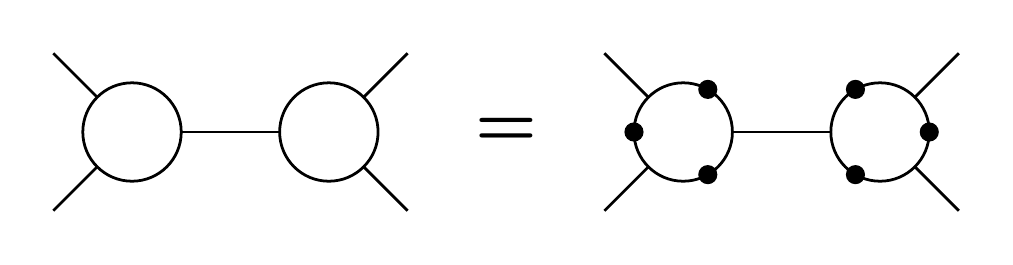} \\
  \vspace{-.5cm}
\caption{A diagrams with independent loops can be reconstructed in one way only from $(n+4)$-pt tree diagrams. It corresponds to nine tree-like diagrams, due to three possible locations of both $\ell_1$ and $\ell_2$.}
\label{2loopcounting2}
\end{figure}

Let us recall fig.~\ref{2loopexample}, where one decomposition of the planar double-box was represented. In fact, other decompositions are possible. If we simply sum over all tree-like diagrams, then we get that each double-box is reconstituted three times, as in fig.~\ref{2loopcounting1}.
On the other hand, diagrams with independent loops can only be reconstituted in one way, as in fig.~\ref{2loopcounting2}.

To understand the multiplicity, let us consider the two vacuum topologies at two loops. 
\begin{figure}
\hspace{3cm}
\begin{tikzpicture}
\node[circle, draw=black, thick, fill=white, minimum size=2cm] at (1,0) {};
\node[circle, draw=black, thick, fill=white, minimum size=1.6cm] at (5,0) {};
\node[circle, draw=black, thick, fill=white, minimum size=1.6cm] at (8,0) {};
\draw[thick] (1,1) -- (1,-1);
\draw[thick] (5.8,0) -- (7.2,0);
\end{tikzpicture}
\caption{Vacuum topologies at two loops.}
\end{figure}
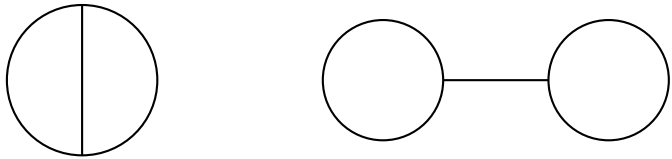
Each two-loop diagram is obtained from one of these by dressing the lines with external legs or trees thereof. There are only 3 cases of interest to us.
\begin{itemize}
\item Case {\bf T1a}: first topology, with the placement of labeled external legs distinguishing the three lines. This is the case of a double-box diagram (planar or non-planar), for instance. The multiplicity is $\rho_{\scalebox{0.6}{T1a}}=6$, with a factor of 3 coming from the consideration in fig.~\ref{2loopcounting1}, and an additional factor of 2 from the exchange of $\ell_1$ and $\ell_2$. 
\item Case {\bf T1b}: first topology, with all external legs and trees thereof attached to only one of the three lines. This makes the other two lines indistinguishable, so that the additional factor of 2 does not apply. The multiplicity is therefore  $\rho_{\scalebox{0.6}{T1b}}=3$.
\item Case {\bf T2}: second topology, of which fig.~\ref{2loopcounting2} provides an example. Notice that at least two external legs (or a single massive tree) must be attached to each loop, otherwise the resulting diagram contains either a tadpole or an external-leg bubble, and should be excluded. The multiplicity is  $\rho_{\scalebox{0.6}{T2}}=2$, given that one may exchange $\ell_1$ and $\ell_2$. 
\end{itemize}

We conclude that the guess \eqref{equ:cubic2loop} is incorrect for generic theories, since it neglects the multiplicity. The scattering-equations version of this guess, in \eqref{equ:2loopgeneric}, is therefore also incorrect.

Again, the exception to this complication are the supersymmetric theories, since all contributions are of the type {\bf T1a}. The multiplicity is therefore an overall normalisation, and the double-forward limit is straightforward.

\subsection{Proposal}\label{sec:finalprop}

We are now in a position to formulate the final proposal. Using the guess \eqref{equ:cubic2loop} as a starting point, the formula should be corrected with the multiplicity coefficients $\rho_\ra$ discussed above, so that the amplitudes are given by
\begin{equation}
\label{equ:cubic2loopfinal}
\boxed{ \quad \mathcal{A}^{(2)}_{\text{YM}} = \int \frac{\d^D\ell_1\,\d^D\ell_2}{\ell^2_1\,\ell^2_2} \sum_{\ra\in \Gamma^{(2)}_{n+4}} \frac{N^{(2)}_\ra\, c^{(2)}_\ra}{\rho_\ra\,D_\ra}\,,
 \qquad \;\;
  \mathcal{A}^{(2)}_{\text{grav}} = \int \frac{\d^D\ell_1\,\d^D\ell_2}{\ell^2_1\,\ell^2_2} \sum_{\ra\in \Gamma^{(2)}_{n+4}} \frac{N^{(2)}_\ra\, \tilde N^{(2)}_\ra}{\rho_\ra\,D_\ra}
  \quad}\,.
\end{equation}
This is the main result of the paper. The formulas exhibit the double-forward limit and the colour-kinematics duality, as intended. The extension to the bi-adjoint scalar theory (or any other theory whose `half-CHY-integrands' can be expanded in terms of Parke-Taylor factors in a similar manner) is straightforward.

With this proposal in hand, we may seek a proof, along the lines of what was achieved at one loop using the factorisation properties of the loop integrand. This is beyond the scope of this paper. At one loop, contributions to the loop integrand that transform homogeneously for the scaling of the loop momentum vanish upon loop integration. This provides a simple check of redundant parts of the loop integrand. At two loops, the analogous argument -- if applied straightforwardly -- is much weaker in identifying expressions that vanish upon loop integration. We hope that a better understanding of this issue may lead to a factorisation-based proof as at one loop.

Another important question concerns the counterpart of \eqref{equ:cubic2loopfinal} in terms of the two-loop scattering equations. We started with the guess \eqref{equ:2loopgeneric}, and even discussed the two `CHY'-integrand prescriptions \eqref{eq:CHY_YM_grav_2loop_cr} and \eqref{eq:CHY_YM_grav_2loop_slash}. In our final proposal, however, we use the CHY-type objects simply to motivate the construction of colour factors and kinematic numerators from their tree-level counterparts via the double-forward limit. The inclusion of the multiplicity factors signals difficulties with the original picture based on the bi-nodal sphere, fig.~\ref{binodalsphere}. We leave this point for section~\ref{sec:degenerations}.

%%%%%%%%%%%%%%%%%%%%%%%%%%%%%%%%%%%%%%%%%%%%%%
%%%%%%%%%%%%%%%%%%%%%%%%%%%%%%%%%%%%%%%%%%%%%%

\section{The Neveu-Schwarz sector of the ambitwistor string}
\label{sec:NSsector}

\begin{figure}[t]
  \hspace{1cm}
\includegraphics[width=13cm]{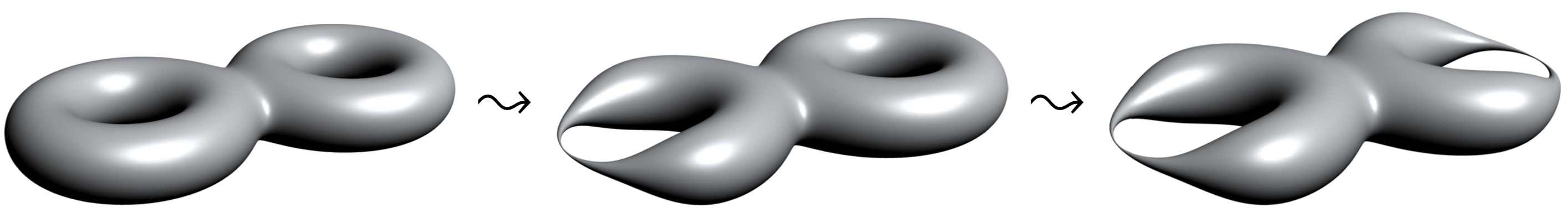} \\
  \vspace{-.5cm}
\caption{A two-step residue argument relates an expression on the genus-two surface to an expression on the bi-nodal sphere.}
\label{doubleresidue}
\end{figure}

In this section, we relate our results to previous work on the calculation of supersymmetric amplitudes from ambitwistor string theory \cite{Geyer:2018xwu}. That calculation is pictured in fig.~\ref{doubleresidue}.
The formulae obtained from ambitwistor string theory at genus two can be turned into much simpler formulae on the bi-nodal sphere, via a residue theorem in moduli space, involving the genus-2 scattering equations. 

The non-supersymmetric kinematic object $\cI_{\text{kin}}^{(2)}$ in \eqref{equ:Ikin2looppf} is based on the double-forward limit. There is, however, another natural candidate for the same object. It arises from the NS sector of the formulae for super-Yang-Mills theory and type II supergravity derived from ambitwistor string theory. This sector contains precisely the states that our new formulae describe. The NS contribution is particularly easy to extract in the RNS formulation of the ambitwistor string. While the RNS formulation  has the disadvantage of not being manifestly supersymmetric, requiring a sum over spin structures (which implements the GSO projection), it is precisely this feature that makes the NS sector easily accessible.

Below, we first review briefly the two-loop integrands of super-Yang-Mills theory and type II supergravity for any number of external particles. We then discuss the projection to pure Yang-Mills theory and NS-NS gravity, and the reason why this is not straightforward, unlike the one-loop case.

\subsection{Supersymmetric two-loop amplitudes}
In the ambitwistor string -- being  a worldsheet model -- two-loop amplitudes arise from correlators on genus-two Riemann surfaces, integrated over the moduli space. As explored in \cite{Geyer:2018xwu}, however, a residue theorem on the moduli space can be used to localise the calculation on the bi-nodal sphere; see fig.~\ref{doubleresidue}. This relies on the localisation provided by the scattering equations in the integration measure, and is thus inherently a field-theory feature that has no straightforward equivalent in conventional string theory. On the bi-nodal Riemann sphere, the correlator then takes the form 
\begin{equation}
 \mathcal{A}^{(2)}_{\text{susy}}=\int \frac{\d^D\ell_1\,\d^D\ell_2}{\ell^2_1\,\ell^2_2} \int_{\mathfrak{M}_{0,n+4}}\!\!\d\mu_{n+4}^{(2,+)}\;\cI^{(2,+)}_{\text{susy}} \,.
 \label{A2loopsusy}
\end{equation}
This is simpler than the naive guess \eqref{equ:2loopgeneric} for generic theories, since only one set of scattering equations is used.\footnote{In the residue theorem, $\alpha=\pm 1$ emerges from a choice between two valid linear combinations for the last modular scattering equation which both lead to a localization on the bi-nodal Riemann sphere; $u_3 = u_{12}+\alpha(u_{11}+u_{22})$ in \cite{Geyer:2018xwu}. This choice then dictates what cross-ratio $\xi^{(\pm)}$ appears in the integrand $\cI_{\text{sugra}}^{(2,\pm)}$.} Of course, the splitting into two parts as in \eqref{equ:2loopgeneric} can be considered, but these should be equivalent up to $\ell_2 \to -\ell_2$. Diagrams of types {\bf T1b} and {\bf T2} do not arise from the particular form of \eqref{A2loopsusy} constructed in \cite{Geyer:2018xwu}, although we will not attempt a general proof here. This would be consistent with the known diagrammatic representations of loop integrands in theories with sufficient amount of supersymmetry. We will take this to mean maximal for super-Yang-Mills theory, and at least half-maximal for supergravity.

Let us review here the expressions for $\cI^{(2,+)}_{\text{susy}}$ constructed in \cite{Geyer:2018xwu}. The moduli-space integrand $\cI^{(2,+)}_{\text{sugra}} $ for type II supergravity was derived directly from the degeneration of the genus-two correlator to the bi-nodal sphere, whereas the analogous super-Yang-Mills integrand $\cI_{\text{SYM}}^{(2,+)}$ was inferred from the double copy and from factorisation requirements. The expressions are
\begin{align}
 \cI_{\text{SYM}}^{(2,+)}= \cI_{\text{susy-kin}}^{(2)}\,\,\cI_{\text{SU}(N_c)}^{(2,\slashed{+})}\,,\qquad \cI_{\text{sugra}}^{(2,+)}=\frac{\sigma_{1^+2^-}\sigma_{2^+1^-}}{\sigma_{1^+1^-}\sigma_{2^+2^-}}\; \cI_{\text{susy-kin}}^{(2)}\,\,\widetilde\cI_{\text{susy-kin}}^{(2)} \,,
 \label{equ:CHYintegrandsSYMsugra}
\end{align}
where the `slash' prescription in the first equation was discussed in \eqref{equ:hatIcol2loopcf},\footnote{We omit the `prime' here, since this is the only prescription for super-Yang-Mills which admits using only one set of scattering equations. The analogous prescription in \eqref{eq:CHY_YM_grav_2loop_slash} would still require contributions from both sets of scattering equations. This can be understood from colour-kinematics duality, and the need to eliminate the contributions from diagrams such as the one on the right-hand side of fig~\ref{jacobi2loop}.} and the cross-ratio in the second equation is the one in \eqref{eq:CHY_YM_grav_2loop_cr}. This cross-ratio stems from the degeneration of the genus-two moduli space to the moduli space of the nodal Riemann sphere. In particular, the non-degenerated genus-two modular parameter maps to a cross-ratio of the nodal points on the sphere, which is itself related to the cross-ratio above. We will see in \cref{sec:cr} that the cross-ratio prescription and the `slash' prescription are equivalent for supergravity.

Important features of the higher-genus Riemann surface are inherited by the  kinematic object $ \cI_{\text{susy-kin}}^{(2)}$. Due to the GSO projection, the genus-two ambitwistor string amplitude contains a sum over spin structures, with the  relative phases determined by modular invariance and unitarity. This sum is of course inherited by the expressions on the nodal Riemann sphere. Combinations of certain terms in this sum have a physical meaning because the spin structures carry the information of what states are propagating in the loops.\footnote{To be precise, there are 16 spin structures, 10 even and 6 odd. Only the even spin structures are relevant for our current purposes. In the conventions of \cite{DHoker:2001kkt,*DHoker:2001qqx,*DHoker:2001foj,*DHoker:2001jaf,*DHoker:2005dys,*DHoker:2005vch}, the spin structures $\delta_{1,2,3,4}$ correspond to NS states propagating in both loops, $\delta_{5,6}$ to a Ramond state in loop 1 (R1), $\delta_{7,8}$ to a Ramond state in loop 2 (R2), and $\delta_{9,0}$ to Ramond states in both loops.} The integrand is thus most conveniently expressed by making this manifest,
\begin{equation}\label{equ:int_nodal_final}
 \mathcal{I}_{\text{susy-kin}}^{(2)}=\mathcal{I}^{\text{NS}}+\mathcal{I}^{\text{R}1}+\mathcal{I}^{\text{R}2}+\mathcal{I}^{\text{RR}}\,,
\end{equation}
where the superscripts NS, R1, R2 and RR denote the states propagating in the loops: 
\begin{center}
\begin{tabular}{c|ccc}
 & \multicolumn{3}{c}{state propagating in} \\
 &  loop 1 &\hspace{5pt}& loop 2 \\\hline
 NS & NS && NS\\
 R1 & R && NS\\
 R2 & NS && R\\
 RR & R && R
\end{tabular}
\end{center}
Here, we take `loop 1' to be the loop containing at least one propagator  involving only $\ell_1$, and similarly for loop 2. For example, the superscript R1 corresponds to a Ramond state propagating in loop 1, but an NS state propagating in loop 2. Thus each term in the sum \eqref{equ:int_nodal_final} contains the contributions from the specified states propagating in the loops. However, it is not possible to extract the NS sector on the  genus-two Riemann surface, since this would violate modular invariance of the integrand. We will see later that this is indicative of a larger issue: while each term in \eqref{equ:int_nodal_final} contains all contributions of the states \emph{to the GSO-sum}, each term by itself is not physically meaningful, because it may lack terms (e.g., involving independent loop momenta) that vanish in the supersymmetric sum. On the bi-nodal Riemann sphere, where the NS-sector can be extracted, the splitting of the GSO-sum thus requires much more care.

Let us take a closer look at the individual terms. Due to their common origin in the ambitwistor string correlator, all terms share the same structure: they are sums over Pfaffians, multiplied by partition functions $\mathcal{Z}_\mathrm{S}$, where S $\in\{$NS, R1, R2, RR$\}$. In particular,
\begin{subequations}\label{equ:integrand_NS-R}
\begin{align}
  \mathcal{I}_n^{\text{NS}}&=4J\,\sum_{n_1,n_2\in\{0,1\}}\,\mathcal{Z}_{\text{NS}}^{(-n_1,-n_2)}\, \pf(M_{\text{NS}})^{(n_1,n_2)}\,,\\
  \mathcal{I}_n^{\text{R}2}&=2J\,\mathcal{Z}_{\text{R}2}^{(0,0)}\, \pf(M_{\text{R}2})^{(0,0)}+2J\,\mathcal{Z}_{\text{R}2}^{(-1,0)}\, \pf(M_{\text{R}2})^{(1,0)}\,,\\
  \mathcal{I}_n^{\text{R}1}&=2J\,\mathcal{Z}_{\text{R}1}^{(0,0)}\, \pf(M_{\text{R}1})^{(0,0)}+2J\,\mathcal{Z}_{\text{R}1}^{(0,-1)}\, \pf(M_{\text{R}1})^{(0,1)}\,,\\
  \mathcal{I}_n^{\text{RR}}&=J\,\mathcal{Z}_{\text{RR}_9}\, \pf(M_{\text{RR}_9})+J\,\mathcal{Z}_{\text{RR}_0}\, \pf(M_{\text{RR}_0})\,.
\end{align}
\end{subequations}
The common prefactor $J$ is a Jacobian from mapping to the moduli space $\mathfrak{M}_{0,n+4}$, and is given by $J=\left(\sigma_{1^+2^+}\sigma_{1^-2^+}\sigma_{1^+2^-}\sigma_{1^-2^-}\right)^{-1}$. The concrete form of the partition functions can be found in the appendix D.2 of \cite{Geyer:2018xwu}. For convenience, we include in our \cref{sec:degen} the explicit coefficients $\mathcal{Z}_{\text{NS}}^{(-n_1,-n_2)}$, since these are required to build up the contribution from NS states, which we will match to new formulae from the previous section. We note a few important features. 
All partition functions depend only on the nodal points $\sigma_{1^\pm}$, $\sigma_{2^\pm}$ and two auxiliary points $x_1$ and $x_2$, related by
\begin{equation}\label{equ:relx1x2RS}
 \omega_{1^+1^-}(x_1)\,\omega_{2^+2^-}(x_2) = \omega_{1^+1^-}(x_2)\,\omega_{2^+2^-}(x_1)\,.
\end{equation}
Here, we have used the notation
\begin{equation}\label{eq:merom-diff}
 \omega_{ij}(\sigma)=\frac{\sigma_{ij}\,\d\sigma}{(\sigma-\sigma_i)(\sigma-\sigma_j)}\,,
\end{equation}
for meromorphic differentials on the sphere, with simple poles at the marked points $\sigma_i$ and $\sigma_j$ of residue $\pm1$. As indicated above, $x_1$ and $x_2$ are auxiliary points, and the  integrands \eqref{equ:integrand_NS-R} are independent of the choice of these points, subject to \eqref{equ:relx1x2RS}.\footnote{In the ambitwistor string, $x_1$ and $x_2$ are insertion points of picture changing operators, so their location  is a choice of gauge. The relation \eqref{equ:relx1x2RS} is a convenient choice simplifying the calculation, and all expressions given here rely on it. }

The superscripts $(\pm n_1,\pm n_2)$ in both the partition functions and the Pfaffians can be thought of as convenient labels. We refer the interested reader to \cite{Geyer:2018xwu}, where their origin from the non-separating degeneration of the genus-two worldsheet is explained.\footnote{The labels correspond to coefficients in the Taylor expansion of each object around $q_1=q_2=0$, where $q_1$ and $q_2$ are the degenerated modular parameters. This is the non-separating degeneration leading to the bi-nodal sphere. Heuristically, we can think of the terms with a $(0,0)$ superscript as the `leading order' contribution for a given state S, and terms with $n_1$, $n_2\neq 0$ as correction terms, coming from $q/q$ contributions in the degeneration, where $q\ll1$ is the modular parameter governing the degeneration. }

\paragraph{The Pfaffian.}
All Pfaffians in \eqref{equ:integrand_NS-R} are defined from the following 
$(2n+2)\times(2n+2)$ matrices $M_{\text{S}}$;
\begin{subequations}
\begin{align}
 & &&M_\text{S}=\begin{pmatrix}A &-C^T\\C&B\end{pmatrix}\,,&&\\
 &A_{x_1x_2}=\px(x_1,x_2)\, S_\text{S}(x_1,x_2)\,,&& A_{x_\beta,j}=P(x_\beta)\cdot k_j \,S_\text{S}(x_\beta,\sigma_j)\,,&& A_{ij}=k_i\cdot k_j \,S_\text{S}(\sigma_i,\sigma_j)\,,\\
 & && C_{x_\beta,j}=P(x_\beta)\cdot \epsilon_j \,S_\text{S}(x_\beta,\sigma_j)\,,&& C_{ij}=\epsilon_i\cdot k_j \,S_\text{S}(\sigma_i,\sigma_j)\,,\\
 & && C_{ii}=-P(\sigma_i)\cdot\epsilon_i\,, && B_{ij}=\epsilon_i\cdot\epsilon_j \, S_\text{S}(\sigma_i,\sigma_j)\,.
\end{align}
\end{subequations}
Note that these matrices do depend on $x_1$ and $x_2$, and only the full integrands $\cI^{\mathrm{S}}$ are independent of the auxiliary points (though not manifestly so). The propagators $S_\mathrm{S}(\sigma_i,\sigma_j)$ depend on the choice S of spin structures, and are detailed in the appendix D.1 of \cite{Geyer:2018xwu}. As $\sigma_i\sim\sigma_j$, they behave as $S_\mathrm{S}(\sigma_i,\sigma_j)\sim \sigma_{ij}^{-1}$ for any S. For convenience, we include in our \cref{sec:degen} the coefficients of $S_\mathrm{NS}(\sigma_i,\sigma_j)$ that will be required to construct $\mathcal{I}_n^{\text{NS}}$ using \eqref{eq:def_higher_order_pf} below. 

The one-form $P(\sigma)$ on the bi-nodal sphere is 
\begin{equation}\label{eq:P}
 P_\mu(\sigma)=\ell_{1\,\mu}\,\omega_{1^+1^-}(\sigma)+\ell_{2\,\mu}\,\omega_{2^+2^-}(\sigma)+\sum_i \frac{k_{i\,\mu}}{\sigma-\sigma_i}\,\d\sigma\,,
\end{equation}
which we saw previously in \eqref{eq:P_2loop}.
In the ambitwistor string, $P$ originates as a field in the worldsheet action. Performing the path integral localises it onto its classical value, given by \cref{eq:P}.
As we saw in \eqref{eq:P2_2loop}, 
the two-loop scattering equations are the conditions that the quadratic differential
\begin{equation}
% \mathfrak{P}_2(\sigma)=
P^2(\sigma)-\ell_1^2\,\omega_{1^+1^-}^2(\sigma)-\ell_2^2\,\omega_{2^+2^-}^2(\sigma) +\left(\ell_1^2+\ell_2^2\right)\omega_{1^+1^-}(\sigma)\,\omega_{2^+2^-}(\sigma)\,
\end{equation}
vanishes everywhere on the Riemann sphere (here again for $\alpha=1$).

Finally, the quantity $\px(x_1,x_2)$, entering the matrices $M_\textrm{S}$ via the component $A_{x_1x_2}$,   is defined as
\begin{equation}
 \px(x_1,x_2) 
 =-\frac{1}{2}\sum_{i,j}\frac{k_i\cdot k_j}{c_1 c_2}\,\Big(c_1\omega_{i*}(x_1)-c_2\omega_{i*}(x_2)\Big)\Big(c_1\omega_{j*}(x_1)-c_2\omega_{j*}(x_2)\Big)\,.
\end{equation}
While this superficially depends on the choice of an auxiliary marked point $\sigma_\ast$, it is easily verified that $\px(x_1,x_2)$ is independent of $\sigma_\ast$ by using the definition of 
the coefficients $c_\beta$ for $\beta=1,2$,
\begin{equation}
 c_\beta = \sqrt{\frac{(x_\beta -\sigma_{1^+})(x_\beta -\sigma_{1^-})(x_\beta -\sigma_{2^+})(x_\beta -\sigma_{2^-})}{\sigma_{1^+1^-}\sigma_{2^+2^-}}}\, \frac1{\d x_\beta}\,.
\end{equation}

\medskip

We are now ready to define the Pfaffian factors in the kinematic integrands $\cI^{\mathrm{S}}$. Let us introduce indices $\mathrm{a},\mathrm{b}=1\dots 2n+2$, which we use to label $\sigma_\mathrm{a}=(x_1,x_2,\sigma_1,\dots,\sigma_n,\sigma_1,\dots,\sigma_n)$ and $v_\mathrm{a}=(P(x_1),P(x_2),k_1,\dots k_n,\epsilon_1,\dots,\epsilon_n)$. For any choice of S, the Pfaffians are then be defined as\footnote{In \cite{Geyer:2018xwu}, a slightly different definition tailored towards the  degeneration of the genus-two worldsheet was presented. It is easily checked that the two definitions are equivalent.}
\begin{subequations}
\begin{align}
 &\pf(M_{\text{S}})^{(0,0)}=\pf\big(M_{\text{S}}\big)\,,   && n_1=n_2=0\\
 &\pf(M_{\text{S}})^{(n_1,n_2)}=\sum_{\mathrm{a}<\mathrm{b}}S_{\mathrm{S}}^{(n_1,n_2)}(\sigma_\mathrm{a},\sigma_\mathrm{b})\, v_\mathrm{a}\cdot v_\mathrm{b}\;%\frac{\partial}{\partial M_{\mathrm{S}\,ab}}\pf\big(M_{\text{S}}\big) 
 {\pf (M_{\mathrm{S}\,}}^{ab}_{ab})\,,&& n_1+n_2>0\,.\label{eq:def_higher_order_pf}
\end{align}
\end{subequations}
The notation ${\pf (M_{\mathrm{S}\,}}^{ab}_{ab})$ indicates that both the rows and columns $\mathrm{a},\mathrm{b}$ have been removed from the matrix $M_{\mathrm{S}}$.

\paragraph{Four particles. } For four external particles, the sum over spin structures can be performed explicitly \cite{Geyer:2018xwu},
giving the integrands
\begin{equation}\label{eq:4pt}
  \cI_{\text{SYM}, n=4}^{(2,+)}= \mathcal{K}\, \widehat{\mathcal{Y}}\,\,\cI_{\text{SU}(N_c)}^{(2,\slashed{+})}\,,\qquad \cI_{\text{sugra},n=4}^{(2,+)}=\big(\mathcal{K}\tilde{\mathcal{K}}\big)\,\widehat{\mathcal{Y}}^2\;\frac{\sigma_{1^+2^-}\sigma_{2^+1^-}}{\sigma_{1^+1^-}\sigma_{2^+2^-}}  \,,
\end{equation}
that is,\, $\cI_{\text{susy-kin}, n=4}^{(2)}=\mathcal{K}\, \widehat{\mathcal{Y}}$. Here, $\mathcal{K}$ is a purely kinematic prefactor that can be extracted,
\begin{align}
 \mathcal{K}=\tr\big(F_1F_2\big)&\tr\big(F_3F_4\big)+\tr\big(F_1F_3\big)\tr\big(F_2F_4\big)+\tr\big(F_1F_4\big)\tr\big(F_2F_3\big)\\
 &-4\,\tr\big(F_1F_2F_3F_4\big)-4\,\tr\big(F_1F_3F_2F_4\big)-4\,\tr\big(F_1F_2F_4F_3\big)\,,\nonumber
\end{align}
where $F_i^{\mu\nu} = k_i^{[\mu}\epsilon_i^{\nu]}$, and similarly $\widetilde{\mathcal{K}} = \mathcal{K}(\epsilon\rightarrow\tilde\epsilon)$. The integrand on the worldsheet then only depends on the marked points via 
\begin{equation}
 \widehat{\mathcal{Y}}=J\mathcal{Y}\,,\qquad \mathcal{Y}=s\Delta_{14}\Delta_{23}-t\Delta_{12}\Delta_{34}\,,
 \label{eq:Yhat}
\end{equation}
where $J=\left(\sigma_{1^+2^+}\sigma_{1^-2^+}\sigma_{1^+2^-}\sigma_{1^-2^-}\right)^{-1}$ is the Jacobian factor defined above, and 
\begin{equation}
 \Delta_{ij}:=\omega_{1^+1^-}(\sigma_i)\omega_{2^+2^-}(\sigma_j)-\omega_{1^+1^-}(\sigma_j)\omega_{2^+2^-}(\sigma_j)\,.
\end{equation}

In the context of the ambitwistor string, the  type II supergravity four-point amplitude was first obtained on the genus-two Riemann surface from  the pure spinor formulation in \cite{Adamo:2015hoa}, following earlier results from the (non-minimal) pure spinor superstring \cite{Berkovits:2006bk,*Mafra:2008ar,*Gomez:2010ad}. After the degeneration to the bi-nodal sphere, the genus-two formula agrees with the one discussed here \cite{Geyer:2016wjx}.

\subsection{A note on the cross-ratio}\label{sec:cr}

In this section, we discuss the equivalence of the cross-ratio and `slash' prescriptions in the case of supergravity. Let us consider eq.~\eqref{equ:CHYintegrandsSYMsugra}. The claim is that
\begin{align}
\cI_{\text{susy-kin}}^{(2,\slashed{+})}=
\frac{\sigma_{1^+2^-}\sigma_{2^+1^-}}{\sigma_{1^+1^-}\sigma_{2^+2^-}}\; \cI_{\text{susy-kin}}^{(2)} \,.
 \label{equ:sugraslashcr}
\end{align}
This imposes a constraint on the kinematic numerators arising from $\cI_{\text{susy-kin}}^{(2)}$.

The four-point check is as follows. Starting from \eqref{eq:4pt}, we can write\footnote{The factor $-1/6$ is introduced to conveniently match the typical normalisation of the two-loop amplitude. Indeed, in section~4.3 of \cite{Geyer:2018xwu}, in particular eq.~(4.17), an overall factor $1/6$ was dropped. This factor is also related to the multiplicity coefficient of the {\bf T1a}-type diagrams of section~\ref{sec:counting} of the present paper.}
\begin{equation}
\cI_{\text{susy-kin}, n=4}^{(2)}=\mathcal{K}\, \widehat{\mathcal{Y}} = -\frac1{6} \sum_{\gamma \in S_{4+2}} \frac{N(1^+, \, \gamma(1,\cdots, 4, 2^+, 2^-)\,,1^-)}{(1^+ \; \gamma(1,\cdots, 4, 2^+, 2^-)\; 1^-)} \,,
\end{equation}
using the notation $(1\,2\,\cdots\,n) \equiv \sigma_{12}\sigma_{23} \cdots \sigma_{n1}$ for simplicity. 
Unlike similar expansions of the non-supersymmetric analogues, this expansion does not rely on the support of scattering equations. The kinematic numerators are defined in the following way. For each $\gamma$ considered in the form $\gamma \equiv \{ A,2^{\pm},B,2^{\mp},C \}$ with $A$, $B$, $C$ non-overlapping subsets of the external particle labels, the corresponding numerator is defined as
\begin{equation}
\mathcal{K}^{-1}\,N(1^+,A,2^{\pm},B,2^{\mp},C,1^-) = s_B \equiv \big( \sum_{i \in B} k_i \big)^2. \label{Ninitial}
\end{equation}
Since the external states are on-shell, momentum conservation dictates that the only non-zero numerators will be those for which $|B|=2$, so that (\ref{Ninitial}) is better expressed as 
\begin{equation}
\mathcal{K}^{-1}\, N(1^+,A,2^{\pm},B,2^{\mp},C,1^-) =
\begin{cases} s_{ij} & B = \{ i,j \} \\ \;\;\; 0 & \mbox{otherwise}
\end{cases}; \qquad i,j \in \{ 1,2,3,4 \}. \label{num}
\end{equation}
To see explicitly that  \eqref{equ:sugraslashcr} holds, one may first check that
\begin{equation}
\sum_{\gamma \in S_4} \frac{N(1^+, \cdots, 2^+, \cdots, 2^-, \cdots, 1^-)}{(1^+ \cdots 2^+ \cdots 2^- \cdots 1^-)} = %\frac{(1^+ 2^-)(2^+ 1^-)}{(1^+ 2^+)(1^- 2^-)}
\frac{\sigma_{1^+2^-}\sigma_{2^+1^-}}{\sigma_{1^+2^+}\sigma_{1^-2^-}}
 \sum_{\gamma \in S_4} \frac{N(1^+, \cdots, 2^-, \cdots, 2^+, \cdots, 1^-)}{(1^+ \cdots 2^- \cdots 2^+ \cdots 1^-)}\,, \label{numrel}
\end{equation}
with the numerators $N$ defined as in (\ref{num}). The sums run over the permutations of the external leg labels. Let us denote the left-hand side of (\ref{numrel}) as $S^+$, the sum of terms containing the relative ordering $(1^+ \cdots 2^+ \cdots 2^- \cdots 1^-)$, and also denote as $S^-$ the analogous sum containing only terms with the relative ordering $(1^+ \cdots 2^- \cdots, 2^+ \cdots 1^-)$. Then (\ref{numrel}) is simply written as
\begin{equation}
S^+ = %\frac{(1^+ 2^-)(2^+ 1^-)}{(1^+ 2^+)(1^- 2^-)} \;
\frac{\sigma_{1^+2^-}\sigma_{2^+1^-}}{\sigma_{1^+2^+}\sigma_{1^-2^-}}\; S^-. \label{numrelS}
\end{equation}
Now the claim above appears quite clearly, since using (\ref{numrelS}) one obtains
\begin{align*}
\frac{\sigma_{1^+2^-}\sigma_{2^+1^-}}{\sigma_{1^+1^-}\sigma_{2^+2^-}} \; \cI_{\text{susy-kin}, n=4}^{(2)}  = 
-\frac{\mathcal{K}}{6}\; \frac{\sigma_{1^+2^-}\sigma_{2^+1^-}}{\sigma_{1^+1^-}\sigma_{2^+2^-}} \; (S^+ + S^-) = -\frac{\mathcal{K}}{6}\; S^+
=  \cI_{\text{susy-kin}, n=4}^{(2,\slashed{+})} \,,
\end{align*}
where, in the second equality, we used
\begin{equation}
\frac{\sigma_{1^+2^-}\sigma_{2^+1^-}}{\sigma_{1^+1^-}\sigma_{2^+2^-}}
\left(1+\frac{\sigma_{1^+2^+}\sigma_{1^-2^-}}{\sigma_{1^+2^-}\sigma_{2^+1^-}}\right) = \xi^{(+)} + \xi^{(-)} =1\,.
\end{equation}

\subsection{The NS sector, degenerations and multiplicity} \label{sec:degenerations}

We will now consider the NS sector, and identify the subtlety that is associated with the multiplicity coefficients in non-supersymmetric theories, from the point of view of the degeneration from genus two to the bi-nodal sphere.

The decomposition of $\cI_{\text{susy-kin}}^{(2)}$ given in \eqref{equ:int_nodal_final} suggests that $\cI^{\mathrm{NS}}$ corresponds to NS states running in both loops. From the perspective of the ambitwistor string, we are thus led to the following (naive) guess for the non-supersymmetric kinematic object:
\begin{equation}\label{equ:int_NS_from_string}
 \mathcal{I}_{\mathbb{A},\text{kin}}^{(2)}=\cI^{\mathrm{NS}}=4J\,\sum_{n_1,n_2\in\{0,1\}}\,\mathcal{Z}_{\text{NS}}^{(-n_1,-n_2)}\, \pf\Big(M_{\text{NS}}^{(n_1,n_2)}\Big)\,.
\end{equation}
Here, we have used the subscript $\mathbb{A}$ (from `ambitwistor string') to distinguish this object from $\mathcal{I}_{\text{kin}}^{(2,+)}$, defined in terms of the double-forward limit in eq.~\eqref{equ:Ikin2looppf}. 

Superficially, the two proposals not only have very different origins, but also take a very different form. $\mathcal{I}_{\text{kin}}^{(2,+)}$ contains only the reduced Pfaffian of the $2(n+4)\times2(n+4)$ matrix $M$, with the state projector implementing the sum over states running in the loops. On the support of the scattering equations, it is invariant under different choices of the reference vector $q$, used to define the projector. Moreover, it depends explicitly on $(\ell_1+\ell_2)^2$, hence the $(+)$ superscript. On the other hand, $\mathcal{I}_{\mathbb{A},\text{kin}}^{(2)}$ is a sum over Pfaffians of $(2n+2)\times(2n+2)$ matrices with partition functions as coefficients. It is invariant under different choices of two additional marked points $x_{1,2}$ on the Riemann sphere, needed in the definitions of both the Pfaffians and the partition functions. Moreover, it only depends explicitly on the the loop momenta via factors $\ell_I\cdot V$, where $V^\mu$ is an external momentum or polarisation vector.

As different as these two objects are, a relation between them is not unprecedented. At one loop, the ambitwistor string correlator gives rise to a similar sum when reduced to the nodal Riemann sphere, which can be shown to match exactly the integrand defined from a forward limit, as  reviewed in \cref{sec:review1loop}.

While we will not give a full proof here, we  have checked numerically for four external particles that the two expressions agree on the support of the $(+)$ scattering equations,
\begin{equation}
\boxed{
 \mathcal{I}_{\mathbb{A},\text{kin}}^{(2)}\stackrel{(+)}{=}\mathcal{I}_{\text{kin}}^{(2,+)}\,.    }
\end{equation}
As discussed above, this is a highly non-trivial check due to the structural differences of the two integrands. Therefore, we expect this equivalence to hold for any number of external particles. This equivalence supports the interpretation of $\cI_{\text{susy-kin}}^{(2)}$ as arising from a double-forward limit, since that corresponds precisely to the construction of $\mathcal{I}_{\text{kin}}^{(2,+)}$. We expect that the other contributions to $\cI_{\text{susy-kin}}^{(2)}$ in \eqref{equ:int_nodal_final} have the same interpretation, if one considers the forward-limit of fermions for the Ramond states.

If the supersymmetric amplitude has a straightforward double-forward limit interpretation, and if its NS part does too, then why does this interpretation fail when applied directly to the non-supersymmetric case, with NS states? We found before that multiplicity coefficients were required in the latter case, distinguishing the classes of diagrams that do not contribute to supersymmetric amplitudes.

\begin{figure}[t]
  \hspace{1cm}
\includegraphics[width=13cm]{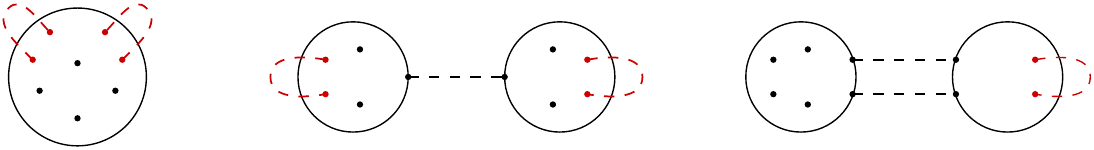} \\
  \vspace{-.5cm}
\caption{Three types of degeneration of a genus-two surface.}
\label{degs}
\end{figure}

We argue that this is due to the failure in the absence of supersymmetry of the straightforward residue derivation depicted in fig.~\ref{doubleresidue}. In the supersymmetric case, as described in detail in \cite{Geyer:2018xwu}, the derivation leads from genus two to a bi-nodal sphere, corresponding to the left figure of fig.~\ref{degs}. However, since the contributions from the individual sectors in \eqref{equ:int_nodal_final} cannot be isolated on the genus-two Riemann surface, we cannot expect the residue derivation to hold for each sector individually. In addition to the terms identified in \eqref{equ:int_nodal_final}, other kinds of degeneration may contribute, which cancel in the supersymmetric sum. For instance, the center and right figures in fig.~\ref{degs} are precisely associated to diagrams of the types {\bf T2} and {\bf T1b}, discussed in section~\ref{sec:counting}, which required distinct multiplicity coefficients with respect to the diagrams of type {\bf T1a} relevant to the supersymmetric case.
We claim that these extra contributions account for the multiplicity coefficients. They could in principle be computed directly, via the `gluing operator' of \cite{Roehrig:2017gbt}, but we will not attempt to do so here, given that the supersymmetric residue argument of \cite{Geyer:2018xwu} was already very intricate.

Therefore, even though our original reasoning was based on the formalism of the scattering equations, our final proposal in section~\ref{sec:finalprop} is expressed only in terms of trivalent tree-like diagrams. We do not discuss further its scattering-equations counterpart, i.e., the explicit correction to the naive guess \eqref{equ:2loopgeneric}. To write a valid proposal based on the scattering equations, there would be further subtleties to deal with, namely those related to the existence of singular solutions to the two-loop scattering equations, for which nodal points coincide. These are irrelevant in the supersymmetric case, but in the non-supersymmetric case their role should be fully understood, as it was at one loop. In contrast to one-loop, however, it is not clear that their contribution remains finite at two loops, and omitting them from the measure would lead to discriminant-style poles. While these integrate to zero at one loop \cite{Cachazo:2015aol}, the analogous statement at two loops remains to be investigated.

To conclude this section, let us emphasise again that the simplifications arising in the maximally supersymmetric case extend to pure half-maximal supergravity. The fact that one of the kinematic objects is supersymmetric is sufficient. Therefore, we can use \eqref{A2loopsusy} with
\begin{align}
\cI_{\text{half-sugra}}^{(2,+)}=\frac{\sigma_{1^+2^-}\sigma_{2^+1^-}}{\sigma_{1^+1^-}\sigma_{2^+2^-}}\; \cI_{\text{susy-kin}}^{(2)}\,\,\widetilde\cI_{\text{kin}}^{(2)} \,.
 \label{equ:CHYintegrandshalfsugra}
\end{align}

%%%%%%%%%%%%%%%%%%%%%%%%%%%%%%%%%%%%%%%%
%%%%%%%%%%%%%%%%%%%%%%%%%%%%%%%%%%%%%%%%

\section{Checks on maximal unitarity cuts}
\label{sec:checks}

As an illustration of our proposal for the two-loop integrand in pure (non-supersymmetric) Yang-Mills theory, we test our formula on unitarity cuts against a known expression. 

We focus on the all-plus helicity configuration at four points. The known expression, originally obtained in \cite{Bern:2000dn}, is
\begin{equation}
\mathcal{A}_4^{(2)}(1^+,2^+,3^+,4^+) = \frac{g^6}{4}\sum_{S_4} \left[C_{1234}^{\text{P}}A_{1234}^{\text{P}} + C_{12;34}^{\text{NP}}A_{12;34}^{\text{NP}} \right] \,,\label{allplus}
\end{equation}
with the colour-ordered amplitudes $A^{\text{P}}$ and $A^{\text{NP}}$ given by
\begin{align}
A_{1234}^{\text{P}} = i\mathcal{T} \bigg\{ sI_4^{\text{P}} &\left[(D_s - 2)(\lambda_1^2 \lambda_2^2 + \lambda_1^2 \lambda_{12}^2 + \lambda_{12}^2 \lambda_2^2) + 16((\lambda_1 \cdot \lambda_2)^2 - \lambda_1^2 \lambda_2^2) \right] (s,t) \nonumber \\
&+ 4(D_s-2)I_4^{\text{bow-tie}} \left[ (\lambda_1^2 + \lambda_2^2)(\lambda_1 \cdot \lambda_2) \right](s) \\
&+ \frac{(D_s-2)^2}{s}I_4^{\text{bow-tie}} \left[\lambda_1^2\lambda_2^2((\ell_1 + \ell_2)^2+s)\right](s,t)\bigg\}\,, \nonumber
\end{align}
\begin{equation}
A_{12;34}^{\text{NP}} = i\mathcal{T} sI_4^{\text{NP}} \left[(D_s - 2)(\lambda_1^2 \lambda_2^2 + \lambda_1^2 \lambda_{12}^2 + \lambda_{12}^2 \lambda_2^2) + 16((\lambda_1 \cdot \lambda_2)^2 - \lambda_1^2 \lambda_2^2) \right] (s,t) \,,
\end{equation}
where $I_4^{\text{P}}[R]$ is a planar double-box with numerator $R$, and likewise $I_4^{\text{NP}}$ denotes a non-planar double-box, and $I_4^{\text{bow-tie}}$ a `bow-tie', as in fig.~\ref{2loopcounting2} but excluding the $1/s$ propagator. Eq.~\eqref{allplus} also includes the corresponding colour factors.
The external kinematics are treated as being four-dimensional, such that the kinematical prefactor $\mathcal{T}$ is defined in terms of the four-dimensional spinor-helicity formalism as
\begin{equation}
\mathcal{T} = \frac{[12][34]}{\langle 12 \rangle \langle 34 \rangle}\,.
\end{equation}
The loop momenta are treated as being in general $D$-dimensional, so that the loop momentum $\ell_I$ has the $(D-4)$-dimensional part $\lambda_I$.

In order to compare this well-known result to our proposal, one may think that we should translate it into the same loop-integrand representation, which for us includes propagators that are not quadratic in the loop momenta, as in \eqref{equ:example2loopprop}. There will be many more terms in the loop integrand. However, if we consider unitarity cuts where both loop momenta are cut, then the propagators match in the two representations, since $\ell_I^2=0$. The contributions to such unitarity cuts should therefore match directly.

For simplicity, we will consider maximal unitarity cuts. We will do this in two cases each for planar and non-planar double-box diagrams. Each case will correspond to a distinct trivalent tree-level-like diagram in our proposal. Its kinematic numerator is constructed from \eqref{equ:num2loops}, considering first the master numerators to be of the type in fig.~\ref{fig:half-ladder-2loop}, and then obtaining the numerator relevant to the cut via Jacobi relations. Since a single trivalent diagram will be considered, we will suppress the multiplicity factor. 

\begin{figure}[t!]
\begin{center}
\includegraphics[scale=0.25]{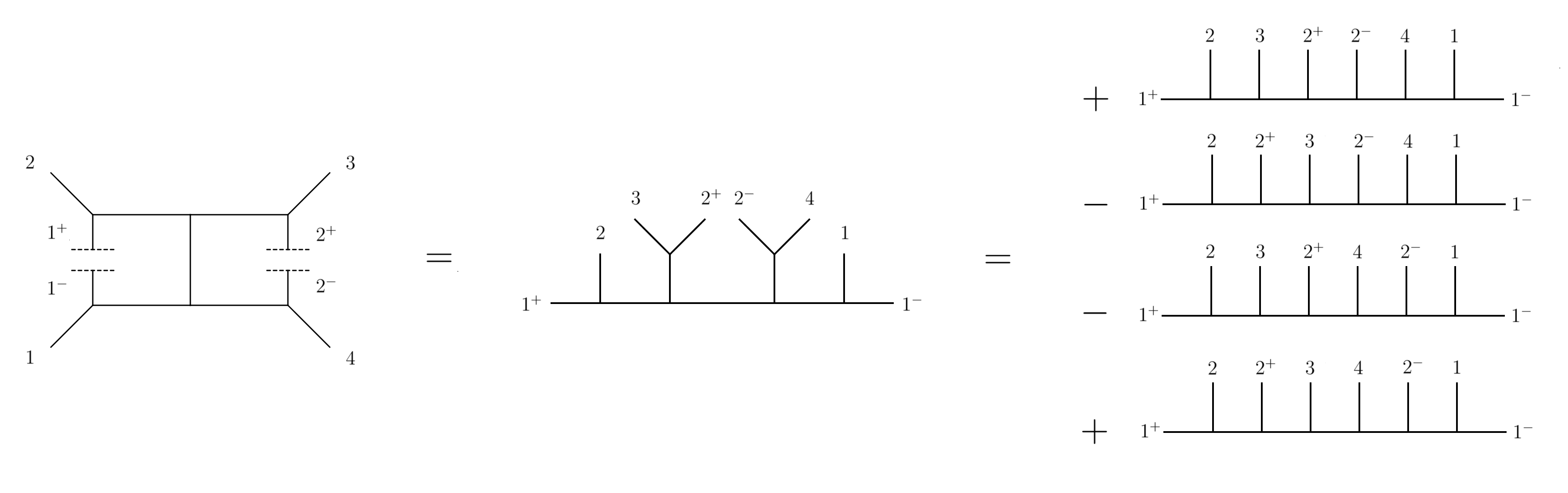}
\end{center}
\caption{Tree-level-like diagram contributing to a planar double-box. Jacobi relations are used twice to derive the kinematic numerator from the numerators of (half-ladder) master diagrams of the type in fig.~\ref{fig:half-ladder-2loop}.}
\label{fig1}
\end{figure}

\begin{figure}[t!]
\begin{center}
\includegraphics[scale=0.25]{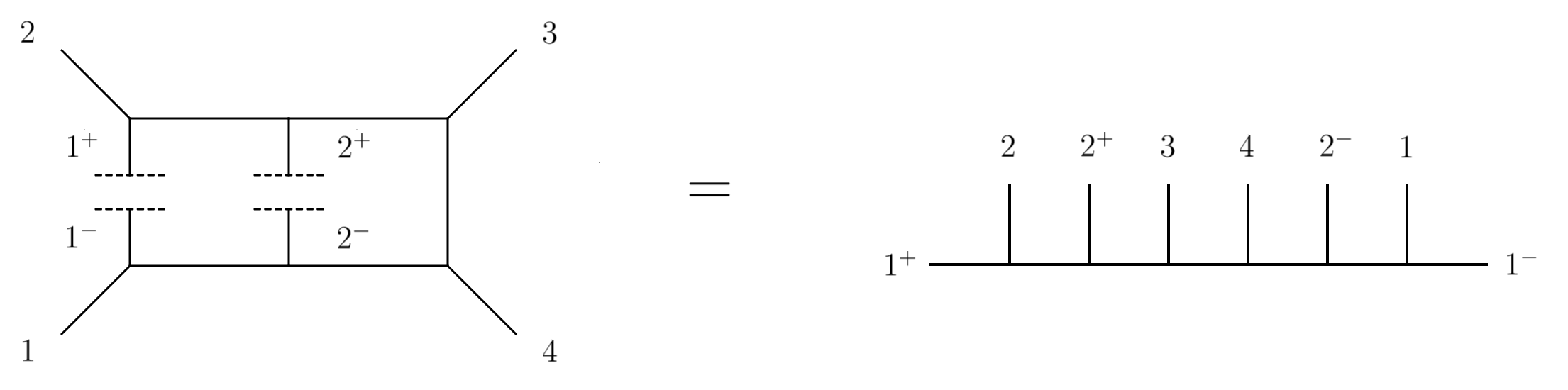}
\end{center}
\caption{Another contribution to a planar double-box, here with a different placement of the loop momenta. In this example, we have a master diagram.}
\label{fig2}
\end{figure}

The two examples of maximal cuts of the planar double-box are shown in figures \ref{fig1} and \ref{fig2}.
Figure \ref{fig1} corresponds to the following term in the integrand,
\begin{equation}
\frac{N^{(2)}(1^+,2\,,[3\,,2^+],[2^-,4\,],1\,,1^-)}{\ell_1^2 \ell_2^2 (\ell_1 + \ell_2 + k_{23})^2(-2\ell_1 \cdot k_1)(2\ell_1 \cdot k_2)(-2\ell_2 \cdot k_4)(2\ell_2 \cdot k_3)} \,,
\label{planbox1}
\end{equation}
where we have used the shorthand notation $k_{ij} \equiv k_i + k_j$, and defined
\begin{equation}
N^{(2)}(\, \cdots, [i,j],\, \cdots) \equiv N^{(2)}(\,\cdots, i\,, j\,,\, \cdots) - N^{(2)}(\,\cdots, j\,, i\,, \,\cdots)\,,
\end{equation}
which conveniently expresses the Jacobi relations.
Note that placing the propagators on-shell gives the same conditions as the maximal cuts for the corresponding Feynman diagram with quadratic propagators:
\begin{equation}
\begin{aligned}
\ell_1^2 =  (\ell_1 + k_2)^2 = (\ell_1 - k_1)^2 =  \ell_2^2 =  (\ell_2 + k_3)^2 = 
(\ell_2 - k_4)^2=  (\ell_1 + \ell_2 + k_{23})^2 = 0. 
\label{plancut1}
\end{aligned}
\end{equation}
We then match -- on the solutions of the maximal cut (\ref{plancut1}) -- the numerator of (\ref{planbox1}) with the numerator of the known result
\begin{equation}
i\mathcal{T}s \left[(D_s - 2)(\lambda_1^2 \lambda_2^2 + \lambda_1^2 \lambda_{12}^2 + \lambda_{12}^2 \lambda_2^2) + 16((\lambda_1 \cdot \lambda_2)^2 - \lambda_1^2 \lambda_2^2) \right] \label{res}
\end{equation}
up to a normalisation numerical factor.\footnote{There is a normalisation convention factor of $i/8$ between the formulas. The factor of $1/8$ can be assigned to explicitly considering the symmetries $\ell_1\leftrightarrow-\ell_1$, $\ell_2\leftrightarrow-\ell_2$ and $\ell_1\leftrightarrow\ell_2$ in our proposal.}

Another maximal cut example for the planar double-box is represented in figure \ref{fig2}, which corresponds to the following term in the integrand,
\begin{equation}\frac{N^{(2)}(1^+,2\,,2^+,3\,,4\,,2^-,1\,,1^-)}{\ell_1^2 \ell_2^2 (-2\ell_1 \cdot k_1)(2\ell_1 \cdot k_2)(\ell_1 + \ell_2 + k_2)^2 (\ell_1 + \ell_2 + k_{23})^2 (\ell_1 + \ell_2 - k_1)^2}\,. \label{pnum2}
\end{equation} 
On the corresponding maximal cut conditions,
\begin{equation}
\begin{aligned}
\ell_1^2 =  (\ell_1 + k_2)^2 = (\ell_1 - k_1)^2 =  \ell_2^2 =  (\ell_1 + \ell_2 + k_{2})^2 = (\ell_1 + \ell_2 + k_{23})^2 =  (\ell_1 + \ell_2 - k_{1})^2 = 0. 
\label{plancut2}
\end{aligned}
\end{equation}
the numerator obtained with our proposal matches that of the known result (\ref{res}).

\begin{figure}[t!]
\begin{center}
\includegraphics[scale=0.25]{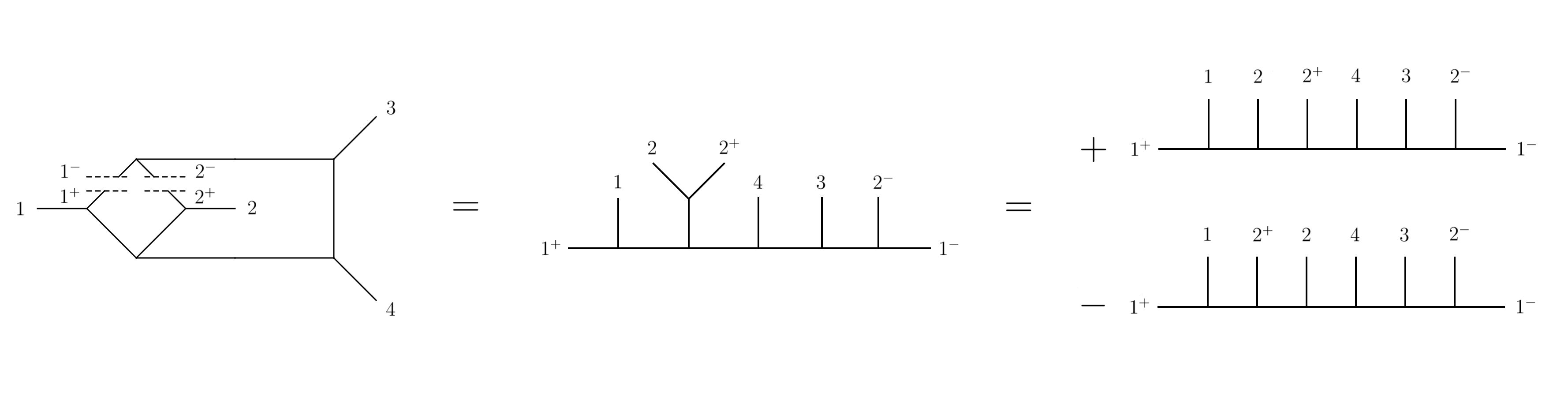}
\end{center}
\caption{Contribution to a non-planar double-box, and corresponding relation master diagrams.}
\label{fig3}
\end{figure}

\begin{figure}[t!]
\begin{center}
\includegraphics[scale=0.25]{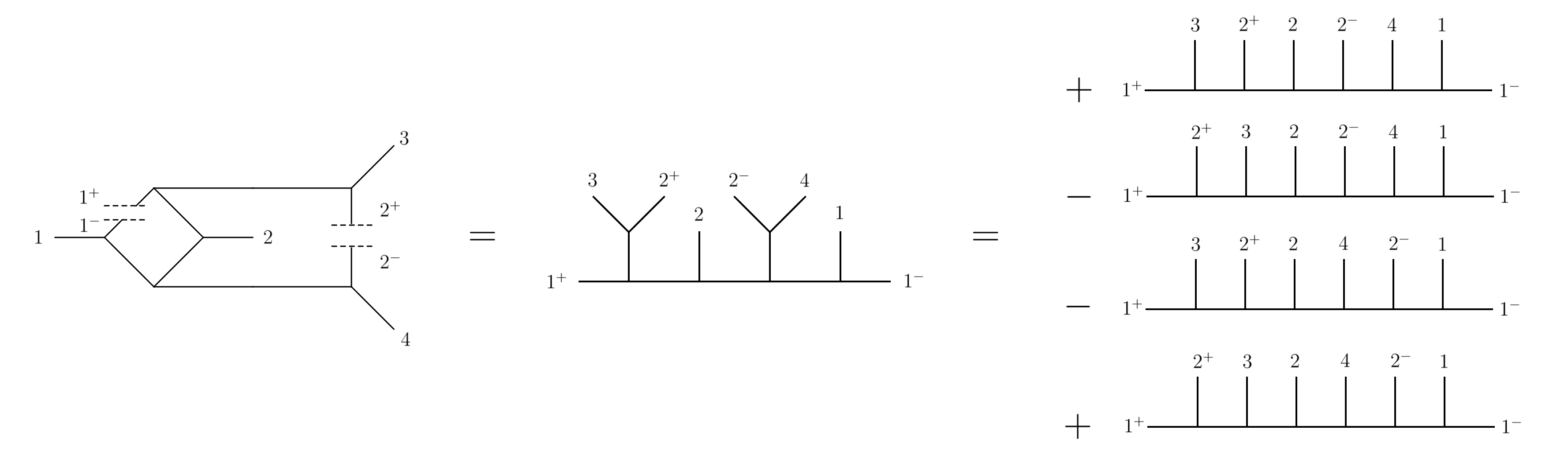}
\end{center}
\caption{Another contribution to a non-planar double-box.}
\label{fig4}
\end{figure}

Now we repeat the procedure with another two cases, this time corresponding to the non-planar double-box. These cases are shown in figures \ref{fig3} and \ref{fig4}.
Figure \ref{fig3} corresponds to the tree-level-like diagram with numerator
\begin{equation}
N^{(2)}(1^+,[2\,,2^+],4\,,4\,,2^-,1^-) \equiv N^{(2)}(1^+,1\,,2\,,2^+,4\,,3\,,2^-,1^-) - N(1^+,1\,,2^+,2\,,4\,,3\,,2^-,1^-)\,, \label{npnum1}
\end{equation}
which, to match with the known result, we evaluate on the corresponding maximal cut solutions:
\begin{equation}
\begin{aligned}
\ell_1^2 =  (\ell_1 + k_1)^2 = \ell_2^2=  (\ell_2 + k_2)^2 = 
(\ell_1 + \ell_2 + k_{12})^2 =  (\ell_1 + \ell_2 - k_3)^2 =  (\ell_1 + \ell_2)^2 = 0. \label{nonplanarcut1}
\end{aligned}
\end{equation}
Similarly, we consider the case in figure \ref{fig4}, where the numerator is
\begin{equation}
\begin{aligned}
N^{(2)}(1^+,[3\,,2^+],2\,,[2^-,4\,],1\,,1^-) \equiv +&N^{(2)}(1^+,3\,,2^+,2\,,2^-,4\,,1\,,1^-) - N^{(2)}(1^+,2^+,3\,,2\,,2^-,4\,,1\,,1^-) \\
- &N^{(2)}(1^+,3\,,2^+,2\,,4\,,2^-,1\,,1^-) + N^{(2)}(1^+,2^+,3\,,2\,,4\,,2^-,1\,,1^-) \label{npnum2}
\end{aligned}
\end{equation} 
and the maximal cut conditions are
\begin{equation}
\begin{aligned}
\ell_1^2 =  \ell_2^2 =  (\ell_2 + k_3)^2 =  (\ell_1 + \ell_2 + k_3)^2 = (\ell_1 + \ell_2 + k_{23})^2 = (\ell_1 - k_1)^2 = (\ell_2 - k_4)^2 = 0. \label{nonplanarcut2}
\end{aligned}
\end{equation}
In each of these cases, the numerators match on the maximal cuts the known result, also given by (\ref{res}).

%%%%%%%%%%%%%%%%%%%%%%%%%%%%%%
%%%%%%%%%%%%%%%%%%%%%%%%%%%%%%

\section{Conclusion}
\label{sec:discussion}

We have presented a construction of two-loop integrands in Yang-Mills theory and gravity based on a double-forward limit of tree-level trivalent diagrams. In this construction, a two-loop version of the colour-kinematics duality follows directly from tree level.

Our work relies on a particular type of loop-integrand representation, which includes non-Feynman propagators. A major open question is how to translate our results into a standard representation, with only Feynman propagators  (whereas translating from the latter representation to the former is straightforward, e.g., using partial fractions and shifts in the loop momenta). Solving this open question could provide a new Feynman-type colour-kinematics prescription at loop level, less restrictive than the original BCJ conjecture \cite{Bern:2010ue}.

We said nothing about loop integration, and yet this is the most glaring open question given the unorthodox type of loop-integrand representation we use, and also given the remarkable recent progress in obtaining analytic results for two-loop amplitudes, e.g., \cite{Badger:2017jhb,*Abreu:2017hqn,*Abreu:2018aqd,*Chicherin:2018yne}. With our current understanding, the direct route of integrating our expressions as they stand seems too hard, given the number of terms involved and, more generally, the fact that integration techniques have been developed for decades with a Feynman-type representation in mind. Even the $i\epsilon$-prescription will differ for our type of representation, as discussed in \cite{Baadsgaard:2015twa}. We can see two options. One is to use our representation to construct the amplitude with unitarity methods. Indeed, given double (or higher) cuts of both loop momenta, the number of contributing terms will be the same as in a Feynman-type representation. The large number of terms is precisely spelling out the cuts. The other option is, of course, to modify our formulae so that they are written in a Feynman-type representation. We hope to report on the latter route in the near future.

%%%%%%%%%%%%%%%%%%%%%%%%%%%%%%
%%%%%%%%%%%%%%%%%%%%%%%%%%%%%%

\section*{Acknowledgements}
We would like to thank Arthur Lipstein, Lionel Mason and Piotr Tourkine for discussions. YG and RM thank the Galileo Galilei Institute for Theoretical Physics and INFN for hospitality and partial support during the workshop ``String Theory from a worldsheet perspective", where part of this work was done. YG gratefully acknowledges support from the National Science Foundation Grant PHY-1606531 and the Association of Members of the Institute for Advanced Study (AMIAS). RM and RS-M are supported by the Royal Society, through a University Research Fellowship and a studentship, respectively.

\appendix

%%%%%%%%%%%%%%%%%%%%%%%%%%%%%%%%%%%%%%%%%%%%%%
%%%%%%%%%%%%%%%%%%%%%%%%%%%%%%%%%%%%%%%%%%%%%%

\section{Two-loop partition functions and propagators on the Riemann sphere}\label{sec:degen}
Below, we list all two-loop partition functions and propagators obtained from even spin structures. All formulas are given on the bi-nodal Riemann sphere. A derivation using the non-separating degeneration of genus-two spin structures can be found in \cite{Geyer:2018xwu}.

To keep the formulae readable, it will be useful to define the cross-ratio
\begin{equation}
 q_3 = \frac{\sigma_{1^+2^+}\sigma_{1^-2^-}}{\sigma_{1^+2^-}\sigma_{1^-2^+}}\,,
\end{equation}
which appears naturally in the double non-separating degeneration from the genus-two Riemann surface as the last modular parameter. Moreover, we will use several other cross-ratios, $v_{1,2}^{\pm}$ and $v^{\pm\pm}$, depending on the nodes $\sigma_{1^{\pm}}$, $\sigma_{2^{\pm}}$ and the marked points $x_{1,2}$, which we will define as needed.

\medskip

We note that all expressions given below simplify considerably when using a convenient gauge choice for the auxiliary marked points $x_1$ and $x_2$. In particular, a nice form of the amplitude can be obtained from $x_1=\sigma_{1^+}$, which automatically implies $x_2=\sigma_{1^-}$  by the identity \eqref{equ:relx1x2RS}. This gauging can be performed most easily by using the 
 parametrization
 \begin{equation}
  x_1 = \sigma_{1^+}+\varepsilon\frac{\sigma_{1^+2^+}\sigma_{1^+2^-}}{\sigma_{2^+2^-}}\,,\qquad x_2=\sigma_{1^-}-\varepsilon\,\frac{\sigma_{1^-2^+}\sigma_{1^-2^-}}{\sigma_{2^+2^-}}\,,
 \end{equation}
and then taking $\varepsilon\rightarrow 0$ in the amplitudes. While this may prove to be a useful tool for working with the two-loop integrand obtained from the ambitwistor string, e.g. in the context of  factorization, we will not need this gauge choice here, and present the following formulas for any choice of $x_1$.

\subsection{Two-loop  partition functions}\label{sec:partition}
 The NS partition functions entering in the kinematic integrand $\cI^{(2)}_{\mathrm{susy-kin}}$ are defined as
\begin{subequations}
 \begin{align}
 & \mathcal{Z}^{(-1,-1)}_{\mathrm{NS}}=\frac{\sqrt{\d x_1 \d x_2}}{x_1-x_2}\frac{q_3^{-2}}{\omega_{1^+1^-}(x_1)\, \omega_{1^+1^-}(x_2)\,\omega_{2^+2^-}(x_1)\,\omega_{2^+2^-}(x_2)}\,,\\
 & \mathcal{Z}^{(-1,0)}_{\mathrm{NS}}\hspace{6pt}=\frac{\sqrt{\d x_1 \d x_2}}{x_1-x_2}\frac{q_3^{-1}}{\omega_{1^+1^-}(x_1)\, \omega_{1^+1^-}(x_2)}Z_8^{(-1,0)}\,,\\
 & \mathcal{Z}^{(0,-1)}_{\mathrm{NS}}\hspace{6pt}=\frac{\sqrt{\d x_1 \d x_2}}{x_1-x_2}\frac{q_3^{-1}}{\omega_{2^+2^-}(x_1)\,\omega_{2^+2^-}(x_2)}Z_8^{(0,-1)}\,,\\
  &\mathcal{Z}^{(0,0)}_{\mathrm{NS}}\hspace{12pt}=10\,q_3\big(1+3q_3+q_3^2\big)\,\mathcal{Z}^{(-1,-1)}_{\mathrm{NS}} + \frac{\sqrt{\d x_1 \d x_2}}{x_1-x_2}\Big(2Z_3^{(-1,0)}Z_3^{(0,-1)}-Z^{(0,0)}\Big)\,,
 \end{align}
\end{subequations}
where the factors of $Z_a^{(-1,0)}$, $Z_a^{(0,-1)}$ and $Z^{(0,0)}$ are given by
 \begin{align*}
  &Z_a^{(-1,0)}=\frac{a}{\omega_{2^+2^-}(x_1)\omega_{2^+2^-}(x_2)}-\frac{\big((x_1-\sigma_{2^+})(x_2-\sigma_{2^+})\,\sigma_{2^-1^+}\sigma_{2^-1^-}-\big(\sigma_{2^+}\leftrightarrow \sigma_{2^-}\big)\big)^2}{\sigma_{2^+2^-}^2\sigma_{1^+2^+}\sigma_{1^+2^-}\sigma_{1^-2^+}\sigma_{1^-2^-}\,\,\d x_1 \d x_2}\,,\\
  &Z_a^{(0,-1)}=\frac{a}{\omega_{1^+1^-}(x_1)\omega_{1^+1^-}(x_2)}-\frac{\big((x_1-\sigma_{1^+})(x_2-\sigma_{1^+})\,\sigma_{1^-2^+}\sigma_{1^-2^-}-\big(\sigma_{1^+}\leftrightarrow \sigma_{1^-}\big)\big)^2}{\sigma_{1^+1^-}^2\sigma_{1^+2^+}\sigma_{1^+2^-}\sigma_{1^-2^+}\sigma_{1^-2^-}\,\,\d x_1 \d x_2}\,,\\
  &Z^{(0,0)}=\frac{\Bigg(\prod_{a=1^+,2^+}\big((x_1-\sigma_{a})(x_2-\sigma_{a})\,\sigma_{1^-2^-}^2\big)^2+\mathrm{perm}(\mathrm{nodes})\Bigg)}{\sigma_{1^+1^-}^2\sigma_{1^+1^-}^2\,\sigma_{1^+2^+}\sigma_{1^+2^-}\sigma_{1^-2^+}\sigma_{1^-2^-}\,\,\d x_1^2 \d x_2^2}\,.
 \end{align*}
 Here, $\mathrm{perm}(\mathrm{nodes})$ denotes the following permutations of the nodal points: $\mathrm{perm}(\mathrm{nodes})=\big(\sigma_{1^+}\leftrightarrow \sigma_{1^-}\big)+\big(\sigma_{2^+}\leftrightarrow \sigma_{2^-}\big)+\big(\sigma_{1^+}\leftrightarrow \sigma_{1^-}, \sigma_{2^+}\leftrightarrow\sigma_{2^-}\big)$. 
 The form degree in the auxiliary insertion points $x_\beta$ cancels out in the final amplitudes formulae. To see this, note that all  partition functions  carry form degree $-3/2$ in  $x_\beta$, whereas the Pfaffians contribute a form degree of  $+3/2$ in each $x_\beta$, coming either from  $A_{x_1x_2}=\px(x_1,x_2)S_\mathrm{S}(x_1,x_2)$ or the product $M_{x_1i}M_{x_2j}\sim S_\mathrm{S}(x_1,\sigma_i)S_\mathrm{S}(x_2,\sigma_j)$.

\subsection{Two-loop propagators}\label{sec:Szego}
The two-loop NS propagators are defined as
\begin{subequations}
\begin{align}
  S_{\text{NS}}(z,w)&=\frac{\sqrt{\d z\,\d w}}{z-w},,\\
  S^{(1,0)}_{\text{NS}}(z,w)&=q_3\frac{\sigma_{1^+1^-}^2\,(z-w) \,\sqrt{\d z\,\d w}}{(z-\sigma_{1^+})(z-\sigma_{1^-})(w-\sigma_{1^+})(w-\sigma_{1^-})}\,,\\
  S^{(0,1)}_{\text{NS}}(z,w)&=q_3\frac{\sigma_{2^+2^-}^2\,(z-w) \,\sqrt{\d z\,\d w}}{(z-\sigma_{2^+})(z-\sigma_{2^-})(w-\sigma_{2^+})(w-\sigma_{2^-})}\,,\\
  S^{(1,1)}_{\text{NS}}(z,w)&=q_3^2\,S^{(1,0)}_{\text{NS}}(z,w)S^{(0,1)}_{\text{NS}}(z,w)\frac{\big((z-\sigma_{1^+})(w-\sigma_{2^+})\,\sigma_{1^-2^-}+(z-\sigma_{2^-})(w-\sigma_{1^-})\,\sigma_{1^+2^+}\big)^2}
  {\sigma_{1^+2^+}\sigma_{1^-2^-}\sigma_{1^+2^-}\sigma_{1^-2^+}(z-w)\,\sqrt{\d z\,\d w}}\,.\nonumber
\end{align}
\end{subequations}

\bibliography{twistor-bib}
\bibliographystyle{JHEP_mod}

\end{document}